\documentclass{emulateapj}

\usepackage{amsmath}

\slugcomment{To appear in ApJ}

\shorttitle{Molecular Gas in the 30 Dor Region of the Large
  Magellanic Cloud} \shortauthors{Pineda, Ott, Klein, Wong, Muller \& Hughes}

\begin{document}

\title{The Influence of Far-Ultraviolet Radiation on the Properties of
  Molecular Clouds in the 30 Dor Region of the Large Magellanic Cloud}

\author{Jorge L. Pineda\altaffilmark{1}}
\affil{Argelander Institut f\"ur Astronomie, Universit\"at Bonn, Auf dem
    H\"ugel 71, D-53121 Bonn, Germany}
\email{Jorge.Pineda@jpl.nasa.gov}
\author{J\"urgen Ott\altaffilmark{2,3}}
\affil{National Radio Astronomy Observatory, P.O. Box O, Socorro, NM 87801}
\author{Ulrich Klein}
\affil{Argelander Institut f\"ur Astronomie, Universit\"at Bonn, Auf dem
    H\"ugel 71, D-53121 Bonn, Germany}

\author{Tony Wong\altaffilmark{4,5}}
\affil{School of Physics, University of New South Wales, Sydney NSW 2052,
Australia}

\author{Erik Muller\altaffilmark{4,6}}
\affil{Department of Astrophysics, Nagoya University, Furo-cho, Chikusa-ku, Nagoya 464-8602, Japan}

\and

\author{Annie Hughes\altaffilmark{4}}

\affil{Centre for Supercomputing and Astrophysics, Swinburne University of Technology, Hawthorn, VIC 3122, Australia}

\altaffiltext{1}{current address: Jet Propulsion Laboratory, California Institute of Technology, 4800 Oak Grove Drive, Pasadena, CA 91109-8099, USA}
\altaffiltext{2}{also at California Institute of Technology 1200 E. California Blvd. Caltech
  Astronomy, 105-24 Pasadena, CA 91125-2400, USA}
\altaffiltext{3}{J\"urgen Ott is a Jansky Fellow of the National Radio
  Astronomy Observatory}
\altaffiltext{4}{also at CSIRO Australia Telescope National Facility, PO Box
  76, Epping NSW 1710, Australia}
\altaffiltext{5}{current address: Department of Astronomy, University of Illinois, Urbana IL 61801, USA}
\altaffiltext{6}{JSPS Fellow}

\begin{abstract}
  We present a complete $^{12}$CO $J = 1 \to 0$ map of the prominent
  molecular ridge in the Large Magellanic Cloud (LMC) obtained with
  the 22-m ATNF Mopra Telescope.  The region stretches southward by
  $\sim$ 2\degr\,(or 1.7\,kpc) from 30~Doradus, the most vigorous
  star-forming region in the Local Group. The location of this
  molecular ridge is unique insofar as it allows us to study the
  properties of molecular gas as a function of the ambient radiation
  field in a low-metallicity environment. We find that the physical
  properties of CO-emitting clumps within the molecular ridge do not
  vary with the strength of the far-ultraviolet (FUV) radiation
  field. Since the peak CO brightness of the clumps shows no
  correlation with the radiation field strength, the observed constant
  value for CO-to-H$_2$ conversion factor along the ridge seems to
  require an increase in the kinetic temperature of the molecular gas
  that is offset by a decrease in the angular filling factor of the CO
  emission. We find that the difference between the CO-to-H$_2$
  conversion factor in the molecular ridge and the outer Milky Way is
  smaller than has been reported by previous studies of the CO
  emission: applying the same cloud identification and analysis
  methods to our CO observations of the LMC molecular ridge and CO
  data from the outer Galaxy survey by \citet{Dame01}, we find that
  the average CO-to-H$_2$ conversion factor in the molecular ridge is
  $X_\mathrm{CO} \simeq (3.9\pm2.5) \times
  10^{20}$~cm$^{-2}$~(K~km~s$^{-1}$)$^{-1}$, approximately twice the
  value that we determine for the outer Galaxy clouds. The mass
  spectrum and the scaling relations between the properties of the CO
  clumps in the molecular ridge are similar, but not identical, to those that have been
  established for Galactic molecular clouds.

\end{abstract}


\section{Introduction}
\label{sec:introduction}

The study of the interstellar medium (ISM) in dwarf galaxies provides
important insights into the star formation history of the Universe.
In massive spiral galaxies the differential rotation causes shear
forces that influence the morphology and evolution of the progenitor
molecular gas and the young stellar component, making the study of the
their relationships difficult.  In general, dwarf galaxies show little
evidence of shear so we can study the relation between star formation
and its prerequisites at their very locations.  They may also be
considered, in a hierarchical galaxy formation scenario, as local
templates of the earliest galaxies. They are thought to be survivors
of the merging processes that lead to the formation of larger
galaxies, such as the Milky Way. Thus, the detailed study of the
low-metallicity molecular gas in dwarf galaxies may help us to
understand the physical conditions under which star formation took
place in the main population of galaxies in the early Universe. This
understanding is crucial in order to interpret observations of the ISM
in galaxies at high redshift.

In dwarf galaxies, stars form out of low-metallicity, low dust-to-gas
ratio dense molecular gas \citep{Lisenfeld98}. Far-ultraviolet (FUV;
6\,eV $<$ $h\nu< 13.6$\,eV) photons from newly formed massive stars
illuminate the progenitor molecular clouds producing photon-dominated
regions (PDRs; e.g. \citealt[][and references
therein]{HollenbachTielens99}). As the FUV field is mainly attenuated
by dust absorption, the penetration of FUV photons is larger in
low-metallicity molecular gas than in its solar metallicity
counterpart.  This produces an enhancement of the photodissociation
rate of CO (increasing the relative abundance of C and C$^+$) while
H$_2$ remains almost unaffected \citep[e.g.][]{vanDishBlack88}.  The
relations between properties of the molecular gas (e.g. mass, CO
luminosity, CO-to-H$_2$ conversion factor) with metallicity and the
strength of the FUV radiation field have been investigated from both
theoretical \citep[e.g.][]{MaloneyBlack88,Elmegreen89,Sakamoto96} and
empirical
\citep[e.g.][]{Dettmar89,Rubio91,Wilson95,Arimoto96,Israel97,Israel00}
perspectives.  The CO-to-H$_2$ conversion factor ($X_{\rm CO} \equiv
N({\rm H}_2)/I_{\rm CO}$; an empirical factor that relates the
intensity of CO [$I_{\rm CO}$] to the column density of H$_2$ [$N({\rm
H}_2)$]) has been found to increase for clouds located in galaxies
with low metallicities and strong FUV radiation fields
\citep{Israel97}. However, $X_\mathrm{CO}$ has also been shown to
decrease with the FUV field by \citet{Weiss01}. This apparent
contradiction arises because $X_\mathrm{CO}$ depends on both the
metallicity {\it and} the FUV radiation field, but mutual variation of
these parameters in different interstellar environments makes it
difficult to determine how they each influence the value of
$X_\mathrm{CO}$\footnote{Note that $X_\mathrm{CO}$ also depends on the
gas density and kinetic temperature \citep{MaloneyBlack88}. However,
the $^{12}$CO $J = 1 \to 0$ line is likely to be thermalized at the
typical densities of molecular clouds and therefore its emission is
not sensitive to the gas density.  The kinetic temperature is related
to the FUV field because the heating in the region where CO is emitted
is dominated by photoelectric heating, which depends on the FUV field
\citep[e.g.][]{BakesTielens94}.}.  Additional complications are
introduced when low-metallicity molecular clouds are observed on
different spatial scales: in particular, $X_\mathrm{CO}$ has been
found to vary with the resolution of the observations
\citep{Rubio93,Verter95}.  At higher angular resolution, the measured
properties of molecular clouds in low metallicity galaxies tend to
approach those of Galactic clouds
\citep{Johansson98,Bolatto03,Leroy2006}.  Ideally, we would like to
conduct high resolution studies of molecular clouds in systems where
one property of the interstellar environment varies while other
environmental conditions remain constant.

The Large Magellanic Cloud (LMC) is an excellent extragalactic
laboratory for such a study.  Its proximity ($\sim$50 kpc;
\citealt[][a value which we will use throughout the paper]{Feast99})
and nearly face-on orientation \citep[$i = 35\degr$;][]{vanderMarel01}
allows us to study individual clouds at high spatial resolution, while
strong spatial variations in the LMC's far-ultraviolet (FUV) radiation
field provide an opportunity to study the influence of the local FUV
field strength on the properties of low-metallicity molecular gas
($30-50\%$ solar; \citealt{Westerlund97}).

The most striking feature in the LMC is the 30\,Doradus H\,{\sc ii}
complex (hereafter 30\,Dor), which is the most massive star-forming
region in the Local Group.  Previous surveys of molecular gas in the
LMC \citep[e.g.][]{Cohen88,Fukui99,Johansson98,Kutner97} have revealed
a prominent complex of molecular clouds extending south from 30\,Dor
for nearly 2~kpc, which we refer to as ``the 30\,Dor molecular
ridge''. Studies of individual clouds in the 30\,Dor molecular ridge
have found gas densities between $\sim$$10^4-10^5$ cm$^{-3}$ and
kinetic temperatures between $\sim$$10-80$\,K
\citep[e.g.][]{Johansson98,Minamidani2007,Pineda2008}.

The young stellar cluster R\,136 located at the center of 30\,Dor produces an
extreme FUV radiation field.  As seen in radio continuum
\citep{Dickel05}, H$\alpha$ \citep{ Gaustad01}, far-infrared
\citep[FIR;][]{Schwering1989} and ultra-violet
\citep[e.g.][]{Smith1987} maps, the strength of the field decreases
with increasing distance from R\,136. The 30\,Dor molecular ridge thus
presents an ideal interstellar environment in which to investigate the
properties of low-metallicity molecular gas as a continuous function
of the ambient FUV radiation field.

Another aspect of interest is the global structure of the
low-metallicity interstellar medium (ISM) of the LMC. Cloud structure
is often characterized by the clump-mass spectrum, which describes how
the total cloud mass is distributed among different-size
structures. \citet{Fukui01} presented a mass spectrum for GMCs in the
LMC, based on NANTEN $^{12}$CO data with spatial resolution of
$\sim$40\,pc. These authors derived a power-law index $\alpha = 1.9$
for the LMC mass spectrum, which is similar to that found in the
Galaxy \citep[1.4 to 1.9;
  e.g.][]{ElmegreenFalgalore96,Heithausen98,Kramer98,Williams94}. Whether
this relation holds at even smaller scales, where clumps have low
masses and are therefore more susceptible to the effect of enhanced CO
photo-dissociation in low-metallicity environments, is still unclear.
This is mainly due to the lack of wide-field, high resolution maps of
$^{12}$CO emission from molecular clouds in low metallicity
environments.

In this paper, we present a complete, fully sampled,  $^{12}$CO $J =
1\to 0$ map of the 30\,Dor molecular ridge in the LMC obtained with the
ATNF Mopra 22-m Telescope.  Parts of this region have been mapped with
the SEST telescope at a similar angular resolution and in the same
transition \citep{Johansson98,Kutner97}.  Cloud properties such as
sizes, CO luminosities, $X_\mathrm{CO}$ factor, etc.  have been
studied in the 30\,Dor and N159/160 regions by \citet{Johansson98} but
not in the southern molecular clouds \citep{Kutner97}, although the
gravitational stability of clumps in this region have been discussed
by \citet{Indebetouw2008}.  We combine our $^{12}$CO data with a map
of the strength of the FUV radiation field, derived from HIRES/IRAS
60\,$\mu$m and 100\,$\mu$m images, in order to study the dependence of
the properties of low-metallicity molecular clouds on the radiation
field. We also present a study of the clump-mass spectrum for the
molecular clouds in the 30\,Dor molecular ridge.  This has not been
attempted in the LMC at the resolution of our observations.  The main
difference between our work and that done with the SEST telescope is
in the analysis of the data. Additionally, the wider coverage of
our map allows us to identify clouds that have not previously been
mapped at this resolution \citep[see also][]{Ott2008}.  The paper is
organized as follows. The observations are summarized in
Sect.~\ref{sec:observations}. The analysis of the data is described in
Sect.~\ref{sec:results} and the results are discussed in
Sect.~\ref{sec:discussion}. A summary is given in
Sect.~\ref{sec:conclusions}.

\section{Observations}
\label{sec:observations}

\begin{figure}[t]
  \includegraphics[width=0.44\textwidth,angle=0]{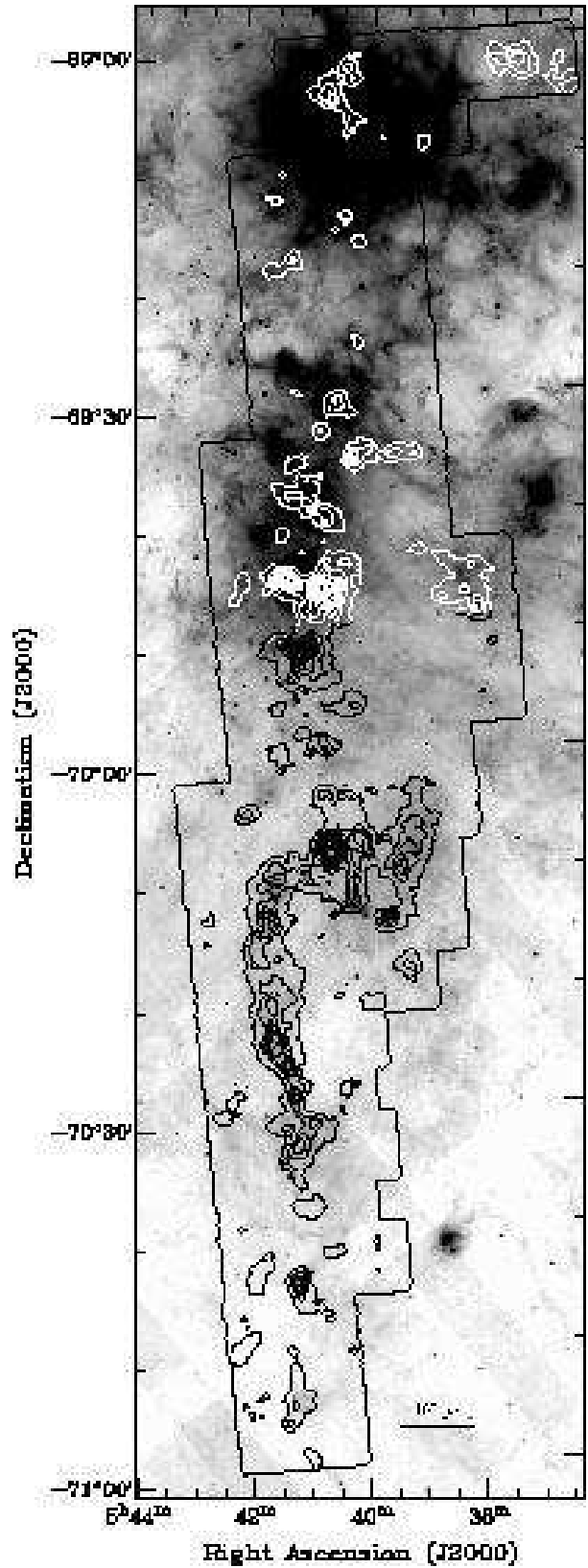}
     \caption{      
       Contour map of the integrated intensity of the $^{12}$CO $J = 1 \to 0$
       line observed toward the 30\,Dor molecular ridge. This contour map is
       overlaid onto an {\it Spitzer} 24\,$\mu$m emission map
       \citep{Meixner06}.  To improve the contrast, we divide the contour map
       into white and black regions.  The contour levels correspond
       to the 5\% to 85\% of the peak intensity (53 K km s$^{-1}$), in steps
       of 10\%.}
\label{fig:ridge}
\end{figure}

The $^{12}$CO $J = 1 \to 0$~(115~GHz) observations were made using the
22-m ATNF Mopra\footnote{The Mopra radio telescope is part of the
Australia Telescope which is funded by the Commonwealth of Australia
for operation as a National Facility managed by CSIRO.} telescope
between June and September 2005. The $^{12}$CO $J =1 \to 0$ integrated
intensity map of the 30\,Dor molecular ridge is shown in
Fig.~\ref{fig:ridge}. The observational results of this map are
presented elsewhere \citep{Ott2008}.

A total of 120 fields, 5\arcmin$\times$5\arcmin~in size, were mapped
using the on-the-fly mapping mode, encompassing the entire 30\,Dor
molecular ridge.  Typical values of the single side-band (SSB) system
temperature $T^{*}_\mathrm{sys}$ of about 600~K were obtained under
clear atmospheric conditions.  Prior to each map, the pointing
accuracy was checked using the SiO maser R~Dor, with corrections
typically below 5\arcsec.

The digital autocorrelator was configured to output 1024 channels
across a 64~MHz (163.8\,km\,s$^{-1}$) band in two polarizations,
corresponding to a velocity resolution of 0.16~km~s$^{-1}$.  We used
the AIPS++ task {\tt LiveData} to calibrate the spectra and to remove
a first-order fit, and another AIPS++ task {\tt Gridzilla} to grid the
data into spectral line cubes.  {\tt LiveData} is the processing
software originally designed for data taken with the Parkes multibeam
receiver and is used to apply system temperature calibration, bandpass
calibration, heliocentric correction, spectral smoothing, and to write
out the data in sdfits format.  {\tt Gridzilla} is a re-gridding
software package that forms three dimensional (RA-Dec-velocity) data
cubes from bandpass-calibrated sdfits files (usually produced by {\tt
  LiveData}).  During the gridding procedure, the data were convolved
with a Gaussian smoothing kernel with a FWHM half of the beam-size
(33\arcsec~at 115~GHz), producing an effective resolution of
45\arcsec~for the data cube.  The spectra were smoothed to an angular
resolution of 60\arcsec~(corresponding to a linear scale of
$\sim$15~pc) and to a velocity resolution of 0.96~km~s$^{-1}$, in
order to increase the signal-to-noise ratio. The Mopra telescope has
an ``inner'' error beam in the range of $40-80$\arcsec, which is
similar to the size of emission. We therefore applied an extended beam
efficiency of $\eta_\mathrm{mb}=0.55$ \citep{Ladd05} to convert from
$T^*_{\rm A}$ to extended beam brightness $(T_{\rm xb})$. The typical
r.m.s. noise of the channel maps is 0.16\,K.

In order to compare the properties of the molecular clouds located in
the 30\,Dor molecular ridge with Galactic clouds of approximately
solar metallicity, we utilize a map of the 2nd Quadrant observed with
the CfA 1.2m telescope and presented by \citet{Dame01}. The 2nd
Quadrant map covers the Galactic plane from {\it l} = 100$\degr$ to
150$\degr$ and {\it b} = -2$\degr$ to 2$\degr$, with a noise level of
about 0.1\,K, at a resolution of 450\arcsec. We use this dataset
(kindly provided by Thomas Dame) because it samples Galactic clouds at
physical scales that are similar to those of our 30\,Dor molecular
ridge map. A further advantage of this data is that blending of
different clouds along each line of sight is negligible for outer
Galaxy clouds.  While the properties of molecular clouds in the
  outer Galaxy are known to be different to those in the inner Galaxy
  \citep[e.g.][]{May1997}, another reason to compare our molecular ridge
  data to the outer Galaxy cloud sample is that molecular clouds in
  the inner Milky Way are subject to conditions that characterize the
  central regions of spiral galaxies (e.g. shear forces, tidal forces,
  strong magnetic fields, cosmic rays), and which may not be common in the LMC.

\section{Analysis and Results}
\label{sec:results}

\subsection{Clump Decomposition}
\label{sec:analysis}

In order to study the properties of the molecular gas in the 30\,Dor
molecular ridge and compare them to properties of clouds in the 
  outer Galaxy, we identify individual clumps in the 30\,Dor
molecular ridge and 2nd Quadrant datasets using the software package
{\tt GAUSSCLUMPS} \citep{StutzkiGuesten90,Kramer98}.  By applying the
same decomposition algorithm to both datasets, we ensure that
differences between the properties of molecular clouds in the two
samples are not due to the decomposition algorithm.  {\tt GAUSSCLUMPS}
decomposes the observed three-dimensional (position and velocity)
intensity distribution into emission from individual clumps by fitting
Gaussians to individual peaks and subtracting them in an iterative
process.  The iterative decomposition process is terminated when the
peak intensities of a new generation of clumps fall below 5 times the
r.m.s. noise of the original datacube. The decomposition process
reports clump parameters such as their position, Local Standard of
Rest (LSR) as velocity, clump orientation, spatial FWHMs along the
spatial axes ($\Delta x$, $\Delta y$), brightness temperature
($T_\mathrm{mb}$), and FWHM linewidths ($\Delta v$). \citet{Kramer98}
described in detail the stability of the results to variations of
input parameters used in the decomposition of several molecular
clouds. For the decomposition of our observations, we adopt the
standard set of stiffness parameters used in their analysis.

As {\tt GAUSSCLUMPS} approximates irregularly-shaped features by
Gaussian clumps, the algorithm can report the identification of small,
low-mass clumps that are the actually the residue of previously
subtracted larger structures. To reduce the impact of this effect on
our analysis, we only accept Gaussian clumps with fitted sizes that
are larger than 110\% of the spatial resolution (i.e. the deconvolved
size of the clump is at least 40\% of the instrumental resolution
along both spatial dimensions).  The FWHM linewidth of a clump is
limited to be larger than 2.44\,km s$^{-1}$ and 1.63\,km s$^{-1}$ for
the 30\,Dor molecular ridge and 2nd Quadrant datasets respectively,
i.e. we reject clumps that occupy a single pixel in velocity space.
At the observed spatial scales, we do not expect that this criterion
excludes a significant number of genuine clumps with small $\Delta
v$. The catalog presented by \citet{Johansson98} has only one clump
with a FWHM linewidth less than 2.44\,km s$^{-1}$, even though their
velocity resolution is $\sim$0.1 km s$^{-1}$. Moreover, only four
clumps with linewidths narrower than 1.63\,km s$^{-1}$ are identified
in the 2nd Quadrant (a similar number is seen in the catalog presented
by \citealt{May1997}).

In order to determine the kinematic distances to the identified clumps
in the 2nd Quadrant dataset, we assume a flat Galactic rotation curve
with rotation velocity $\Theta_\mathrm{LSR}=220$ km\,s$^{-1}$ and
radius $R_\sun$ = 8.5\,kpc. We discarded all (194) clumps with kinematic
distances lower than 2\,kpc, as the estimation of kinematic
distances is not accurate enough for such local emission.

We identify a total of 67 and 118 clumps in the 30\,Dor molecular
ridge and 2nd Quadrant datasets, respectively.          The average
(deconvolved) radius of the identified clumps in the 30\,Dor molecular
ridge and 2nd Quadrant is $\sim$10\,pc, with a dispersion of $\sim35$\%.

 \begin{figure}[t]
           \centering 
           \includegraphics[width=0.5\textwidth,angle=0]{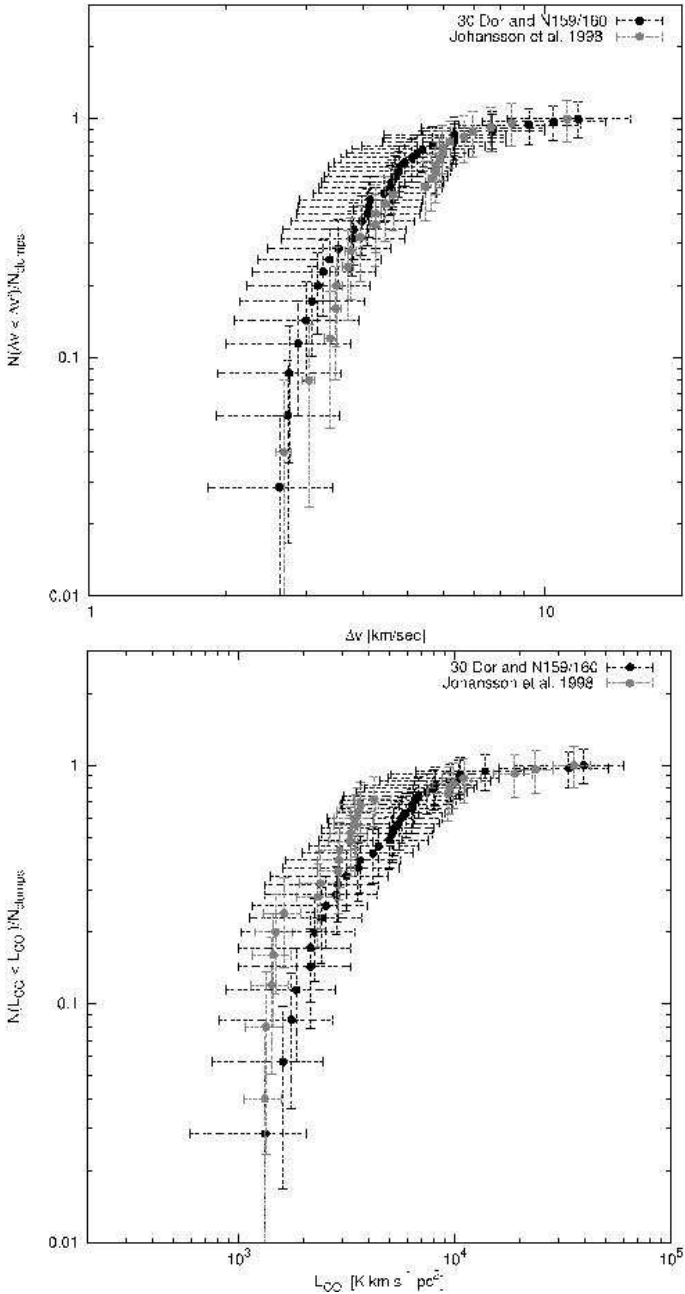}
      \caption{ Cumulative histograms of the FWHM linewidth and CO luminosity
        resulting from our decomposition of the 30\,Dor ridge and the
        data published by \citet{Johansson98}.  The FWHM linewidth is
        not deconvolved by the instrumental resolution.  For all curves, the horizontal error bars reflect
       the uncertainties in the observed properties and the vertical
       error bars represent $\sqrt N$ counting errors.  }
 \label{fig:johansson_heyer}
 \end{figure}

 The derived properties of the identified clumps in the 30\,Dor
molecular ridge are listed in Table\,1.  The CO luminosity of a clump
is estimated using

\begin{align}  & \frac{L_\mathrm{CO}}{\mathrm{K\,km\,s^{-1}\,pc^2}} =  \nonumber \\  & \left (
\frac{2\pi}{8\log(2)}
\right )^{3/2}
\left (
\frac{\Delta x}{\mathrm{pc}}
\right )
\left (
\frac{\Delta y}{\mathrm{pc}}
\right )
\left (
\frac{\Delta v}{\mathrm{km\,s^{-1}}}
\right )
\left (
\frac{T_\mathrm{mb}}{\mathrm{K}}
\right ).
\end{align}

We derive the virial mass, $M_{\mathrm{vir}}$, using

\begin{equation}
\label{eq:1}
\frac{M_\mathrm{vir}}{\mathrm{M}_\odot} = k \,\left(  \frac{R}{\mathrm{pc}}\right ) \, \left(\frac{\Delta v'}{\mathrm{km\,s^{-1}}} \right)^2,
\end{equation}

where R is the deconvolved clump radius and $\Delta v'$ is the deconvolved
FWHM of the observed line profile.  The factor $k$ depends on the assumed
density profile of a clump \citep{Maclaren88}. In the following, we adopt a
value of $k = 210$, corresponding to a spherical cloud with uniform density.
Different values of $k$ (i.e. different density profiles) only change the
scaling but not the slope of any relation involving $M_\mathrm{vir}$.  The
deconvolved clump radius is defined as

\begin{equation}
\label{eq:2} 
\frac{R}{\mathrm{pc}} = 
\frac{1}{2}
\left ( 
\frac{\Delta x'}{\mathrm{pc}}
 \right)^{1/2} 
\left ( 
\frac{\Delta y'}{\mathrm{pc}}
 \right)^{1/2}, 
\end{equation}

where $\Delta x'$ and $\Delta y'$ are the projected deconvolved FWHM
angular sizes of the clump.  We choose to use deconvolved values in
the calculation of the virial mass because it is an intrinsic property
of the cloud. Some published studies do not deconvolve the cloud
parameters, and instead quote cloud property measurements that are
convolved with the instrumental resolution.  This leads to an
overestimate of the virial mass, in particular if the cloud size and
FWHM linewidth are of the same order as the instrumental resolution.
For example, the values of $X_\mathrm{CO}$ obtained in our survey (see
Section \ref{sec:molec-cloud-prop}) are $\sim40$\% larger if we use
undeconvolved values for the clump radius and FWHM linewidth; for the
2nd Quadrant data, using undeconvolved values increases the derived
$X_\mathrm{CO}$ values by $\sim30$\%. The different methods used to
calculate the virial mass and other cloud parameters complicates a
quantitative comparison between our results and those of previous
studies.

 We estimate the value of $X_\mathrm{CO}$ for clumps in the 30\,Dor
 molecular ridge assuming that they are in virial equilibrium.  In
 this case, $X_\mathrm{CO}$ is calculated dividing the clump's virial
 mass - which we assume is an accurate measure of the total mass of
 H$_2$ present - by its CO luminosity.  These quantities are the
 result of multiplying the peak H$_2$ column density and CO intensity
 ($I_{\rm CO}$) of a (Gaussian) clump by its area. Therefore, $X_{\rm
   CO} \equiv N({\rm H}_2)/I_{\rm CO}=M_{\rm vir}/L_{\rm CO}$. Note
 that $L_{\rm CO}$ does not need to be calculated from deconvolved
 values of $R$ and $\Delta v$ since $L_{\rm CO}$ is independent of the
 resolution of the observations.

The effect of adopting different values of $k$ in eq.~(\ref{eq:1}) is
to reduce $X_\mathrm{CO}$ by 10\% and 66\%, for density profiles
of the form $n(r)\propto r^{-1}$ and $n(r)\propto r^{-2}$,
respectively. However, the relative values of $X_\mathrm{CO}$ in the
30\,Dor ridge and the 2nd Quadrant do not depend on the adopted
density profile, as their virial masses were calculated with the same
formula.

The errors in the clump properties are estimated by considering the
r.m.s.  noise in the positions where clumps were identified and by
assuming that the clump parameters obtained by {\tt GAUSSCLUMPS}
($\Delta x$, $\Delta y$, $\Delta v$) have errors of 30\%. We use error
propagation to obtain uncertainties for $X_\mathrm{CO}$,
$M_\mathrm{vir}$, and $L_\mathrm{CO}$.  The virial masses and CO
luminosities have typically errors of 50\%, whereas the error of
$X_\mathrm{CO}$ is typically 70\%.

We test our decomposition by comparing our measurements of the
properties of clumps in the 30\,Dor molecular ridge to those listed in
Table~1 of \citet{Johansson98}. \citet{Johansson98} derive cloud
parameters by fitting circular Gaussians to each intensity peak in
their integrated intensity maps, so an exact correspondence between
the results of the two studies is not to be expected. In
Figure~\ref{fig:johansson_heyer}, we show cumulative histograms of the
FWHM linewidths\footnote{ Note that we use undeconvolved values of the
FWHM linewidth for this comparison, in order to be consistent with
\citet{Johansson98}. As explained in
Section~\ref{sec:molec-cloud-prop}, we use deconvolved values of the
clump parameters in our analysis. } and the CO luminosities of clumps
identified by us and by \citet{Johansson98}. We exclude clumps in the
30\,Dor molecular ridge dataset that are located outside the 30\,Dor
and N159/160 regions observed by \citet{Johansson98}. Furthermore, we
only consider clumps identified by \citet{Johansson98} that have
similar CO luminosities to the clumps in the 30\,Dor molecular ridge,
as their observations are more sensitive than ours. Note that we
choose $\Delta v$ and $L_\mathrm{CO}$ because they are independent of
angular resolution which is different in the 30\,Dor ridge and
\citet{Johansson98} datasets.  For our decomposition, the mean
($\pm$standard deviation) $L_\mathrm{CO}$ is (6.9$\pm$7.9)$\times
10^3$\,K\,km\,s$^{-1}$\,pc$^{2}$ and $\Delta v$ is (4.9$\pm$2.1)
km\,s$^{-1}$.  In case of \citet{Johansson98}, the mean ($\pm$standard
deviation) $L_\mathrm{CO}$ is (6.6$\pm$8.0)$\times
10^3$\,K\,km\,s$^{-1}$\,pc$^{2}$ and $\Delta v$ is (5.3$\pm$1.9
km\,s$^{-1}$).  Kolmogorov-Smirnov (KS) tests between the
distributions of $\Delta v$ and $L_\mathrm{CO}$ for the two datasets
indicate that the probabilities (P) that the cloud samples are drawn
from the same parent distribution are 0.2, and 0.1, respectively (We
consider P$<0.05$ to be statistically significant and P$<0.1$ to be
marginally significant).  The $L_{\rm CO}$ and $\Delta v$ measurements
for the \citet{Johansson98} and 30\,Dor ridge datasets cover a similar
range of values, even though differences in decomposition methods used
by \citet{Johansson98} and us almost certainly affect the exact
distribution of the clump properties.

In the following, we compare the results from our decomposition
of the 30\,Dor molecular ridge dataset with those of the decomposition of the 2nd
Quadrant data. As noted above, this ensures that any difference (or
similarity) between these two datasets is not an effect of the adopted
decomposition method.

\begin{deluxetable*}{ccrrrrrcccrccccccc}
\tabletypesize{\tiny}\tablecaption{Clump properties in the 30\,Dor molecular ridge}
 \tablewidth{0pt}
 \tablehead{
 \colhead{Clump} 
 &\colhead{$\alpha$(J2000)} 
 &\colhead{$\delta$(J2000)} 
 &\colhead{$v_\mathrm{lsr}$}
 &\colhead{$\Delta$$v'^a$} 
 &\colhead{$R^a$} 
 &\colhead{$T_\mathrm{mb}$}
 &\colhead{log($L_\mathrm{CO})^{b}$} 
 &\colhead{log($M_\mathrm{vir})^{b}$} 
 &\colhead{$X_\mathrm{CO}$} 
 &\colhead{$\chi^{c}$} \\
 \colhead{\#} 
 &\colhead{[h\,m\,s]} 
 &\colhead{[\degr\,\arcmin\,\arcsec]} 
 &\colhead{[km\,s$^{-1}$]}
 &\colhead{[km\,s$^{-1}$]} 
 &\colhead{[pc]} 
 &\colhead{[K]}
 &\colhead{[K km s$^{-1}$]} 
 &\colhead{[M$_\sun$]} 
 &\colhead{$10^{20}$[cm$^{-2}$/(K km s$^{-1}$)]} &\colhead{[$\chi_0$]} 
 }
\startdata
1 & 05:36:14.455 & -69:01:48.77 & 272.1 &7.6$\pm$2.3 & 7.6$\pm$1.6       & 1.3$\pm$0.2 & 3.7 & 5.0 & 8.1$\pm$5.8     &5     \\  
2 & 05:38:30.816 & -69:02:04.52 & 249.6 &4.6$\pm$1.4 & 8.7$\pm$1.8       & 1.6$\pm$0.1 & 3.7 & 4.6 & 3.6$\pm$2.5     &119 \\
3 & 05:35:56.511 & -69:02:17.78 & 259.4 &5.3$\pm$1.6 & 10.8$\pm$2.3      & 2.3$\pm$0.1 & 4.0 & 4.8 & 2.8$\pm$2.0     &6     \\ 
4 & 05:38:50.957 & -69:03:52.51 & 246.7 &6.3$\pm$1.9 & 8.4$\pm$1.8       & 1.6$\pm$0.1 & 3.8 & 4.8 & 5.0$\pm$3.5     &407 \\
5 & 05:35:04.678 & -69:04:44.26 & 256.4 &2.9$\pm$0.9 & 10.8$\pm$2.3      & 2.2$\pm$0.2 & 3.7 & 4.3 & 1.6$\pm$1.1     &8     \\ 
6 & 05:38:37.524 & -69:07:28.57 & 245.7 &2.7$\pm$0.8 & 12.0$\pm$2.5      & 1.0$\pm$0.1 & 3.5 & 4.3 & 3.0$\pm$2.1     &318 \\
7 & 05:39:49.663 & -69:12:51.66 & 244.8 &3.0$\pm$0.9 & 7.2$\pm$1.5       & 1.1$\pm$0.1 & 3.3 & 4.1 & 3.5$\pm$2.5     &22   \\ 
8 & 05:38:57.857 & -69:15:40.54 & 248.7 &3.7$\pm$1.1 & 9.8$\pm$2.1       & 0.8$\pm$0.1 & 3.4 & 4.4 & 5.3$\pm$3.8     &36   \\ 
9 & 05:40:17.065 & -69:17:26.83 & 282.8 &2.8$\pm$0.8 & 8.8$\pm$1.9       & 0.8$\pm$0.1 & 3.2 & 4.2 & 4.3$\pm$3.1     &7    \\ 
10 & 05:39:46.005 & -69:29:03.82 & 224.3 &2.6$\pm$0.8 & 9.3$\pm$2.0      & 0.9$\pm$0.1 & 3.2 & 4.1 & 3.3$\pm$2.4     &17  \\ 
11 & 05:39:04.934 & -69:29:52.53 & 257.4 &3.9$\pm$1.2 & 7.8$\pm$1.7      & 2.0$\pm$0.1 & 3.6 & 4.4 & 2.5$\pm$1.8     &81  \\
12 & 05:38:55.865 & -69:34:40.62 & 247.7 &10.4$\pm$3.1& 8.4$\pm$1.8      & 1.6$\pm$0.2 & 4.0 & 5.3 & 8.5$\pm$6.0     &22  \\
13 & 05:38:16.886 & -69:34:52.60 & 262.3 &4.8$\pm$1.4 & 9.4$\pm$2.0      & 1.6$\pm$0.2 & 3.8 & 4.7 & 3.3$\pm$2.4     &12  \\
14 & 05:39:55.512 & -69:35:15.59 & 228.2 &4.0$\pm$1.2 & 7.4$\pm$1.6      & 1.2$\pm$0.2 & 3.4 & 4.4 & 4.5$\pm$3.2     &21  \\
15 & 05:39:28.026 & -69:36:16.22 & 241.8 &5.0$\pm$1.5 & 6.9$\pm$1.5      & 0.8$\pm$0.1 & 3.4 & 4.6 & 7.6$\pm$5.5     &32  \\
16 & 05:39:48.753 & -69:37:27.78 & 229.1 &3.2$\pm$1.0 & 14.8$\pm$3.1     & 1.5$\pm$0.2 & 3.8 & 4.5 & 2.3$\pm$1.6     &76  \\
17 & 05:39:28.185 & -69:40:16.23 & 241.8 &4.7$\pm$1.4 & 5.9$\pm$1.3      & 1.0$\pm$0.1 & 3.3 & 4.4 & 5.8$\pm$4.1     &51  \\
18 & 05:39:25.996 & -69:43:16.27 & 237.9 &3.1$\pm$0.9 & 10.8$\pm$2.3     & 1.5$\pm$0.1 & 3.7 & 4.3 & 2.0$\pm$1.4     &33  \\
19 & 05:37:28.212 & -69:44:27.83 & 235.0 &3.7$\pm$1.1 & 20.5$\pm$4.3     & 1.0$\pm$0.1 & 4.0 & 4.8 & 2.8$\pm$2.0     &10  \\
20 & 05:40:12.264 & -69:44:39.09 & 233.1 &6.3$\pm$1.9 & 8.7$\pm$1.8      & 3.3$\pm$0.1 & 4.1 & 4.9 & 2.4$\pm$1.7     &109 \\
21 & 05:38:56.010 & -69:44:40.63 & 232.1 &11.8$\pm$3.5& 8.7$\pm$1.8      & 0.9$\pm$0.1 & 3.8 & 5.4 & 17.4$\pm$12     &11   \\
22 & 05:39:36.495 & -69:45:40.07 & 236.9 &6.3$\pm$1.9 & 15.0$\pm$3.2     & 4.6$\pm$0.2 & 4.6 & 5.1 & 1.4$\pm$1.0     &173 \\
23 & 05:37:39.654 & -69:46:52.09 & 237.9 &4.0$\pm$1.2 & 9.8$\pm$2.1      & 1.4$\pm$0.1 & 3.6 & 4.5 & 3.6$\pm$2.6     &5   \\
24 & 05:39:28.457 & -69:47:04.23 & 233.1 &4.0$\pm$1.2 & 11.6$\pm$2.5     & 2.1$\pm$0.1 & 3.9 & 4.6 & 2.3$\pm$1.6     &47  \\
25 & 05:39:56.251 & -69:47:15.59 & 233.1 &2.4$\pm$0.7 & 7.4$\pm$1.6      & 1.5$\pm$0.2 & 3.3 & 4.0 & 2.0$\pm$1.4     &49  \\
26 & 05:39:42.389 & -69:47:51.94 & 239.9 &5.6$\pm$1.7 & 9.9$\pm$2.1      & 1.8$\pm$0.2 & 3.9 & 4.8 & 4.1$\pm$2.9     &36  \\
27 & 05:37:04.846 & -69:48:03.17 & 251.6 &4.3$\pm$1.3 & 8.1$\pm$1.7      & 2.2$\pm$0.1 & 3.8 & 4.5 & 2.5$\pm$1.8     &10  \\
28 & 05:39:40.131 & -69:49:04.00 & 254.5 &9.2$\pm$2.8 & 11.0$\pm$2.3     & 1.0$\pm$0.2 & 3.9 & 5.3 & 11.1$\pm$8.1    &13  \\
29 & 05:39:59.945 & -69:50:39.49 & 235.9 &7.6$\pm$2.3 & 12.8$\pm$2.7     & 4.1$\pm$0.2 & 4.5 & 5.2 & 2.1$\pm$1.5     &10  \\
30 & 05:40:33.628 & -69:50:50.31 & 236.9 &5.2$\pm$1.6 & 8.8$\pm$1.9      & 0.8$\pm$0.1 & 3.5 & 4.7 & 7.9$\pm$5.7     &12  \\
31 & 05:39:51.973 & -69:53:15.71 & 237.9 &4.5$\pm$1.3 & 7.2$\pm$1.5      & 1.2$\pm$0.2 & 3.5 & 4.5 & 4.6$\pm$3.4     &5       \\
32 & 05:40:27.061 & -69:55:26.58 & 237.9 &3.4$\pm$1.0 & 10.9$\pm$2.3     & 1.1$\pm$0.2 & 3.6 & 4.4 & 3.4$\pm$2.5     &3       \\
33 & 05:39:13.648 & -69:55:28.46 & 234.0 &4.5$\pm$1.3 & 10.8$\pm$2.3     & 1.5$\pm$0.2 & 3.8 & 4.7 & 3.7$\pm$2.7     &5       \\
34 & 05:39:47.587 & -69:58:03.83 & 238.9 &2.6$\pm$0.8 & 5.4$\pm$1.1      & 1.2$\pm$0.2 & 3.1 & 3.9 & 2.6$\pm$1.9     &4       \\
35 & 05:40:29.749 & -69:59:26.48 & 235.0 &4.7$\pm$1.4 & 9.8$\pm$2.1      & 1.0$\pm$0.2 & 3.6 & 4.7 & 5.7$\pm$4.1     &2       \\
36 & 05:39:50.201 & -70:02:51.77 & 231.1 &5.4$\pm$1.6 & 7.4$\pm$1.6      & 1.5$\pm$0.2 & 3.6 & 4.7 & 4.9$\pm$3.5     &2       \\
37 & 05:41:10.078 & -70:04:12.62 & 227.2 &4.0$\pm$1.2 & 6.3$\pm$1.3      & 1.5$\pm$0.1 & 3.5 & 4.3 & 3.4$\pm$2.4     &1       \\
38 & 05:38:06.973 & -70:05:52.53 & 248.7 &3.9$\pm$1.2 & 7.5$\pm$1.6      & 0.9$\pm$0.2 & 3.3 & 4.4 & 5.4$\pm$3.9     &5       \\
39 & 05:38:25.773 & -70:05:52.67 & 239.9 &4.4$\pm$1.3 & 18.4$\pm$3.9     & 1.8$\pm$0.2 & 4.2 & 4.9 & 2.0$\pm$1.4     &5       \\
40 & 05:39:17.514 & -70:07:04.41 & 236.0 &6.2$\pm$1.9 & 10.1$\pm$2.1     & 2.8$\pm$0.1 & 4.1 & 4.9 & 2.8$\pm$2.0     &3       \\
41 & 05:39:49.299 & -70:07:33.79 & 230.2 &6.9$\pm$2.1 & 11.1$\pm$2.4     & 3.8$\pm$0.2 & 4.4 & 5.0 & 2.2$\pm$1.6     &3       \\
42 & 05:39:34.007 & -70:07:40.13 & 239.9 &3.6$\pm$1.1 & 12.4$\pm$2.6     & 1.4$\pm$0.1 & 3.7 & 4.5 & 2.7$\pm$1.9     &3       \\
43 & 05:39:44.609 & -70:07:51.91 & 241.8 &4.3$\pm$1.3 & 8.4$\pm$1.8      & 2.8$\pm$0.2 & 3.9 & 4.5 & 1.9$\pm$1.3     &4       \\
44 & 05:39:27.011 & -70:09:16.26 & 229.2 &4.9$\pm$1.5 & 8.0$\pm$1.7      & 1.4$\pm$0.1 & 3.6 & 4.6 & 4.6$\pm$3.3     &3       \\
45 & 05:38:39.881 & -70:09:52.69 & 233.1 &4.8$\pm$1.4 & 9.3$\pm$2.0      & 1.8$\pm$0.1 & 3.8 & 4.6 & 3.1$\pm$2.2     &4       \\
46 & 05:40:42.497 & -70:10:01.97 & 227.2 &4.3$\pm$1.3 & 17.4$\pm$3.7     & 2.5$\pm$0.2 & 4.3 & 4.8 & 1.6$\pm$1.1     &5       \\
47 & 05:41:01.407 & -70:10:25.09 & 220.4 &2.4$\pm$0.7 & 12.3$\pm$2.6     & 1.2$\pm$0.2 & 3.5 & 4.2 & 2.2$\pm$1.6     &2       \\
48 & 05:39:29.459 & -70:11:28.22 & 224.3 &5.4$\pm$1.6 & 11.1$\pm$2.4     & 2.1$\pm$0.1 & 4.1 & 4.8 & 2.8$\pm$2.0     &6       \\
49 & 05:38:37.524 & -70:11:52.69 & 246.7 &4.0$\pm$1.2 & 12.4$\pm$2.6     & 1.0$\pm$0.2 & 3.6 & 4.6 & 4.5$\pm$3.3     &3       \\
50 & 05:40:51.138 & -70:13:37.60 & 225.2 &5.4$\pm$1.6 & 11.0$\pm$2.3     & 2.5$\pm$0.1 & 4.1 & 4.8 & 2.6$\pm$1.8     &3       \\
51 & 05:38:51.720 & -70:14:16.66 & 237.9 &3.1$\pm$0.9 & 13.2$\pm$2.8     & 0.9$\pm$0.1 & 3.6 & 4.4 & 3.3$\pm$2.4     &3       \\
52 & 05:41:06.653 & -70:14:48.84 & 222.3 &3.6$\pm$1.1 & 15.3$\pm$3.2     & 2.2$\pm$0.2 & 4.1 & 4.6 & 1.4$\pm$1.0     &2       \\
53 & 05:40:40.816 & -70:16:50.07 & 225.2 &2.8$\pm$0.8 & 17.4$\pm$3.7     & 1.0$\pm$0.2 & 3.7 & 4.5 & 2.5$\pm$1.8     &2       \\
54 & 05:38:42.269 & -70:17:40.69 & 218.4 &2.5$\pm$0.8 & 9.1$\pm$1.9      & 1.9$\pm$0.2 & 3.5 & 4.1 & 1.6$\pm$1.1     &2       \\
55 & 05:40:49.285 & -70:18:25.70 & 229.1 &4.1$\pm$1.2 & 18.4$\pm$3.9     & 1.8$\pm$0.2 & 4.2 & 4.8 & 2.1$\pm$1.5     &2       \\
56 & 05:39:15.584 & -70:20:52.45 & 219.4 &6.4$\pm$1.9 & 8.8$\pm$1.9      & 1.1$\pm$0.1 & 3.7 & 4.9 & 7.4$\pm$5.3     &1       \\
57 & 05:40:58.143 & -70:23:13.30 & 230.1 &5.0$\pm$1.5 & 19.1$\pm$4.1     & 3.5$\pm$0.2 & 4.6 & 5.0 &  1.2$\pm$0.8    &4      \\
58 & 05:39:44.489 & -70:27:27.94 & 227.2 &5.2$\pm$1.6 & 9.3$\pm$2.0      & 1.1$\pm$0.2 & 3.6 & 4.7 & 5.7$\pm$4.1     &1       \\
59 & 05:42:05.959 & -70:29:33.28 & 230.1 &4.3$\pm$1.3 & 10.0$\pm$2.1     & 1.1$\pm$0.2 & 3.6 & 4.6 & 4.5$\pm$3.3     &0.3     \\
60 & 05:40:27.827 & -70:30:38.64 & 227.2 &5.6$\pm$1.7 & 9.8$\pm$2.1      & 1.9$\pm$0.2 & 3.9 & 4.8 & 3.8$\pm$2.7     &2       \\
61 & 05:40:39.896 & -70:31:26.16 & 245.7 &3.0$\pm$0.9 & 6.6$\pm$1.4      & 1.4$\pm$0.2 & 3.3 & 4.1 & 2.8$\pm$2.0     &2       \\
62 & 05:40:32.831 & -70:32:50.45 & 224.3 &2.8$\pm$0.8 & 10.6$\pm$2.2     & 0.8$\pm$0.1 & 3.3 & 4.2 & 3.7$\pm$2.7     &3       \\
63 & 05:40:45.094 & -70:35:13.96 & 230.1 &2.9$\pm$0.9 & 7.0$\pm$1.5      & 1.0$\pm$0.2 & 3.2 & 4.1 & 3.7$\pm$2.7     &2       \\
64 & 05:41:31.859 & -70:42:23.61 & 230.1 &5.6$\pm$1.7 & 11.9$\pm$2.5     & 1.0$\pm$0.1 & 3.8 & 4.9 & 5.7$\pm$4.1     &0.3     \\
65 & 05:40:53.211 & -70:43:13.65 & 235.0 &2.5$\pm$0.8 & 9.1$\pm$1.9      & 1.2$\pm$0.2 & 3.4 & 4.1 & 2.4$\pm$1.7     &2       \\
66 & 05:40:55.748 & -70:44:13.54 & 238.9 &5.0$\pm$1.5 & 8.1$\pm$1.7      & 2.9$\pm$0.2 & 3.9 & 4.6 & 2.3$\pm$1.6     &2       \\
67 & 05:41:11.555 & -70:54:00.85 & 234.0 &4.1$\pm$1.2 & 11.8$\pm$2.5     & 1.4$\pm$0.1 & 3.9 & 4.6 & 2.7$\pm$1.9     &1       
\enddata
\tablenotetext{a}{Deconvolved by the instrumental resolution.}
\tablenotetext{b}{The CO luminosities and virial masses have typically errors of 50\%. }
\tablenotetext{c}{Strength of the FUV radiation field in units of the \citet{Draine78} field.  They have typical  errors   of 20\%, as this is the accuracy of the HIRAS/IRAS images used to derive this quantity (see text).   }
\end{deluxetable*}

\subsection{Estimation of the Far-ultraviolet Radiation Field}
\label{sec:vari-far-ultr}

\begin{figure}[t]
   \centering
   \includegraphics[width=0.65\textwidth,angle=0]{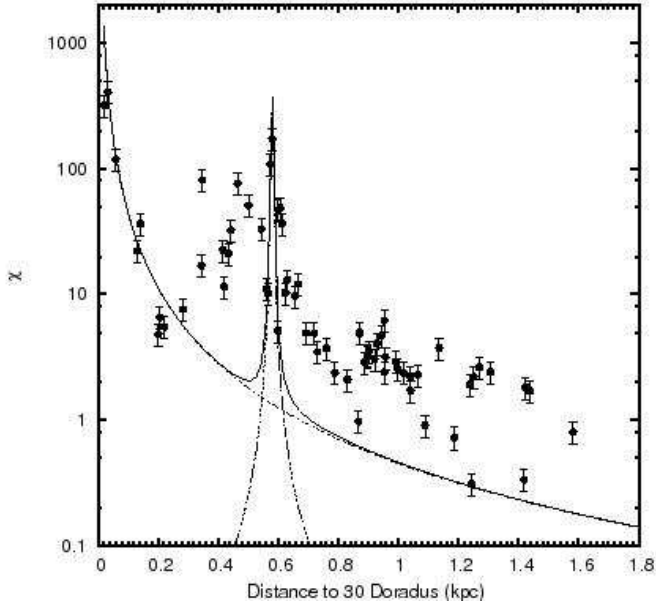}
      \caption{Strength of the FUV radiation field as a function of
        distance to 30\,Dor. The values of $\chi$ are in units of the
        \citealt{Draine78} field. Each data-point represents the value
        of $\chi$ at the central position of each identified
        clump. The main and secondary peaks corresponds to clumps
        close to the 30\,Dor and N159 regions.  The lines represents
        the geometrical dilution of the FUV field in case 30\,Dor and
        N159 are considered the only sources of FUV radiation (see text).}
\label{fig:chi}
 \end{figure}

An important feature of the 30\,Dor molecular ridge is that the
low-metallicity molecular gas is exposed to very different strengths
of the FUV radiation field. This can be seen in maps of different
tracers of the FUV radiation field, e.g. at radio continuum
\citep{Dickel05}, H$\alpha$ \citep{ Gaustad01}, far-infrared
\citep[][]{Schwering1989}, and ultra-violet \citep[e.g.][]{Smith1987}
wavelengths.  For all tracers, the intensity peaks in regions that are
closely associated with star formation, and declines in locations that
are more distant from regions of active star formation.  Assuming that
metallicity gradients are negligible in the 30\,Dor molecular ridge
\citep{Dufour1984}, we can therefore study clump properties as a
function of the strength of the FUV radiation field.  To derive a map
of the strength of the FUV radiation field in the 30\,Dor molecular
ridge, we follow the method presented in \citet{Mookerjea2006} and
\citet{Kramer2008}.  A map of the FIR continuum intensity
($I_\mathrm{FIR}$) is constructed using HIRES/IRAS 100\,$\mu$m and
60\,$\mu$m images obtained from the IPAC data center.  The resolution
of these images has been enhanced to about 110\arcsec and 72\arcsec,
respectively using a maximum correlation method
\citep{Aumann1990}. The 60\,$\mu$m image is then convolved to
110\arcsec, in order to match the resolution of the 100\,$\mu$m image.
To estimate $I_\mathrm{FIR}$ between 42.5\,$\mu$m and 122.5\,$\mu$m,
we use the relationship (\citealt{Helou1988}, see also \citealt{Nakagawa1998}):
 
\begin{align}  & \frac{I_\mathrm{FIR}}{\mathrm{erg\,cm^{-2}\,s^{-1}\,sr^{-1}}} =  \nonumber \\  & 3.25\times10^{-11} 
\left [ 
\frac{f_{60\mu \mathrm{m}}}{\mathrm{Jy\,sr^{-1}}} 
\right ]
+
1.26\times10^{-11} 
\left [ 
\frac{f_{100\mu \mathrm{m}}}{\mathrm{Jy\,sr^{-1}}} 
\right ].
\end{align}

The dust grains in molecular clouds absorb most of the incident FUV
photons, which are later re-radiated in the FIR continuum.  Assuming
that the contribution from other heating mechanisms such as cosmic
rays can be neglected, the strength of the FUV field can be derived
from $I_\mathrm{FIR}$ according to

\begin{equation}
\frac{\chi}{\chi_0}
=
\frac{1}{1.71}
\frac{4\pi}{(2\cdot1.6\times10^{-3})}
\frac{I_\mathrm{FIR}}{\mathrm{erg\,cm^{-2}\,s^{-1}\,sr^{-1}}}.
\end{equation}

Here, $\chi$ is expressed in units of the \citet{Draine78} field
($\chi_0$), which is related to the \citet{Habing68} field
(1.6$\times$10$^{-3}$ erg s$^{-1}$ cm$^{-2}$) by a factor of 1.71 when
averaged over the full FUV spectrum.  It is likely that the
contribution of FUV photons with $h\nu < 6$ eV increases the dust
heating by a factor of $\sim2$ \citep{TielensHollenbach85}, and that
the bolometric dust continuum intensity is a factor of $\sim$2 larger
than $I_\mathrm{FIR}$ \citep{Dale2001}: these two corrections
approximately cancel out.  Note that in the case where molecular
clouds do not fill the beam, this method for estimating the strength
of the FUV field provides only a lower limit.

Typical angular sizes of clumps are between $\sim$66\arcsec\,to
200\arcsec, and as the resolution of the $\chi$ map is about
110\arcsec, its value represents an average over an entire clump.  The
calibration of the $\chi$ map appears to be consistent with previous
estimates of the FUV radiation field in the vicinity of the 30\,Dor
molecular ridge. Near the N159W cloud (Clump \#22) for example, we
obtain $\chi$=173, which is in good agreement with an independent
determination from UV, radio continuum, and IRAS 60\,$\mu$m emission
($\chi = 120-350$) made by \citet{Israel96}. Our estimate for $\chi$
in N159W is also consistent with the results of PDR models for this
molecular cloud \citep[e.g.][]{Pak1998,Pineda2008}. Although the
$\chi$=173 may be considered low for a massive star-forming region (in
the Orion Bar, for example, \citealt{Walmsley00} and
\citealt{YoungOwl00} find $\chi \simeq 10000$), \citet{Israel96} argue
that the low metallicity and low dust-to-gas ratio of this region
increase the FUV photon free path lengths, resulting in a greater
geometric dilution of the FUV radiation field. We stress, however,
that for this analysis we are interested in the dependence of the CO
clumps parameters on {\it relative} variations of the FUV radiation
field.

In Fig.~\ref{fig:chi}, we show $\chi$ as a function of the distance to
30\,Dor. Each point corresponds to the value of $\chi$ in the central
position of each molecular clump identified in the 30\,Dor molecular
ridge. The value of $\chi$ for each clump is also listed in Table~1.
The strength of the FUV field peaks in clouds near the 30\,Dor and
N159 H\,{\sc ii} regions, and then decreases for more distant
positions. The thick line represents the expected distribution of
$\chi$ if 30\,Dor and N159 were the only sources of FUV radiation,
where we have used the value of $\chi$ at the clump closest to the
center of each H\,{\sc ii} region to represent the radiation source,
and considering only geometrical dilution of the FUV field.  The
individual contribution from each of these regions is shown by a
dashed line. Fig.~\ref{fig:chi} shows that the radiation from 30\,Dor
and N159 is not sufficient to explain the observed distribution of
$\chi$, and that additional OB stars along the 30\,Dor ridge must
therefore contribute to the local FUV radiation field.

\subsection{Clump properties as a function of the FUV radiation field}
\label{sec:nh2ico-conv-fact}

\begin{figure}[t]
  \centering \includegraphics[width=0.5\textwidth,angle=0]{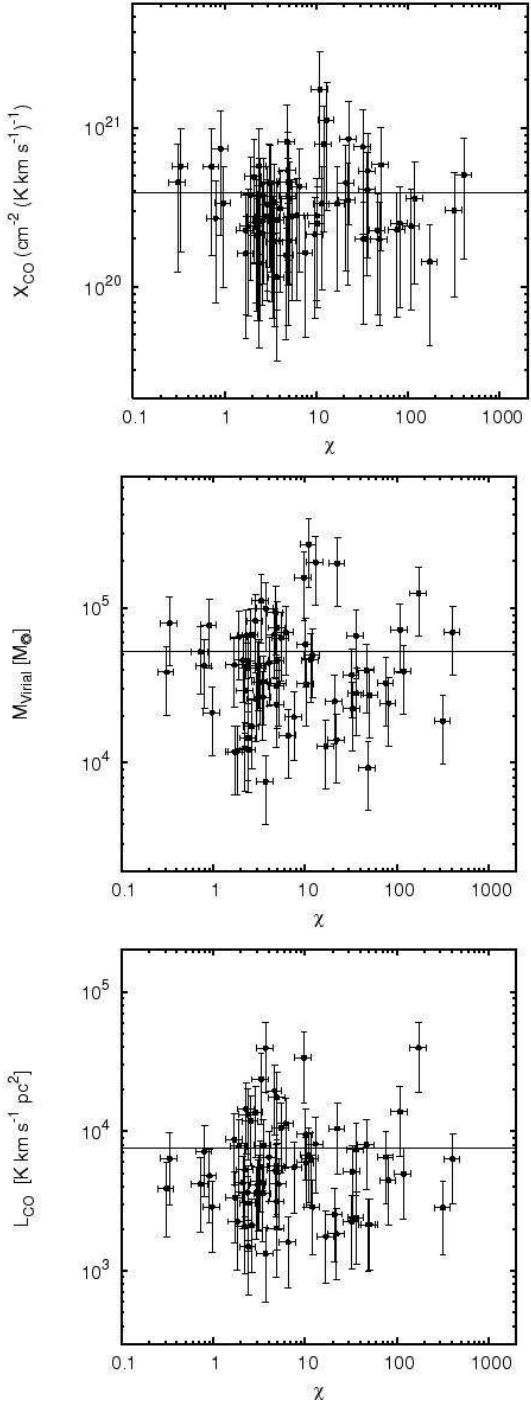}
     \caption{The CO-to-H$_2$ conversion factor ($X_\mathrm{CO}$), the virial
       mass ($M_\mathrm{vir}$), and the CO luminosity ($L_\mathrm{CO}$) as
       functions of the strength of the FUV radiation field. The horizontal lines represent
 the mean values of the data (see Table\,2). }
\label{fig:xco}
\end{figure}

\begin{figure}[t]
          \centering \includegraphics[width=0.5\textwidth,angle=0]{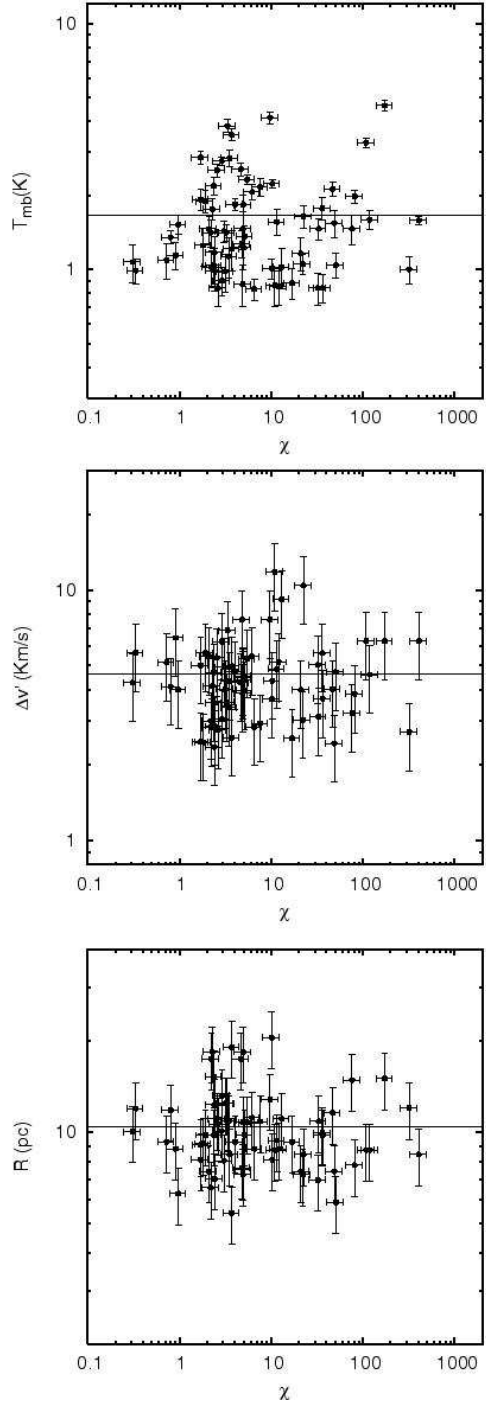}
     \caption{The brightness temperature ($T_\mathrm{mb}$), clump
       FWHM linewidth ($\Delta v'$), and clump radius ($R$) as functions
       of the strength of the FUV radiation field. 
The horizontal lines represent
 the mean values of the data (see Table\,2)}
\label{fig:tmb_dv_r}
\end{figure}

Figure~\ref{fig:xco} shows $L_\mathrm{CO}$, $M_\mathrm{vir}$, and
$X_\mathrm{CO}$ plotted as a function of the FUV radiation field.  The
horizontal lines represent the mean values of the data given in
Table\,2. As we can see, these quantities do not show any correlation
over three  orders of magnitude in $\chi$.  To interpret the plots
  of $M_\mathrm{vir}$ and $L_\mathrm{CO}$ with $\chi$, we show the
  brightness temperature ($T_\mathrm{mb}$), clump FWHM linewidth
  ($\Delta v'$), and clump radius ($R$) as a function of $\chi$ in
  Figure~\ref{fig:tmb_dv_r}.  Again, the horizontal lines represent
  the mean values of the data given in Table\,2.  These basic clump
  parameters also appear to be independent of the strength of the FUV
  radiation field. In Section~\ref{sec:discussion}, we discuss
  possible reasons why the properties of the CO clumps in the 30\,Dor
  molecular ridge are insensitive to the strength of the FUV field.  

\subsection{Comparison with clouds in the  outer Galaxy}
\label{sec:molec-cloud-prop}

\begin{figure}[t]
       \centering
       \includegraphics[width=0.42\textwidth,angle=0]{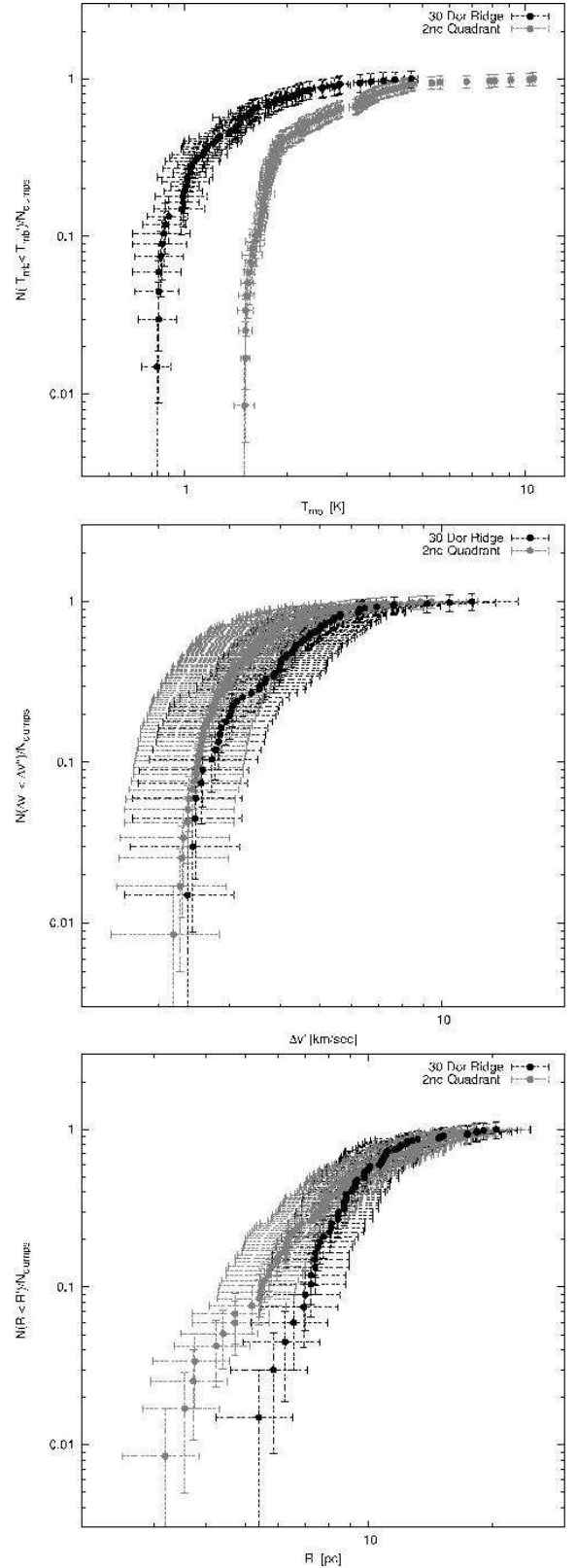}
        \caption{Cumulative histograms of the brightness temperature
       ($T_\mathrm{mb}$), clump FWHM linewidth ($\Delta v'$), and
       clump radius ($R$) derived in the 30\,Dor Ridge and 2nd
       Quadrant.  For all curves, the horizontal error bars reflect
       the uncertainties in the observed properties and the vertical
       error bars represent $\sqrt N$ counting errors.
       }\label{fig:hist_basic}
\end{figure}

\begin{figure}[t]
       \centering \includegraphics[width=0.41\textwidth,angle=0]{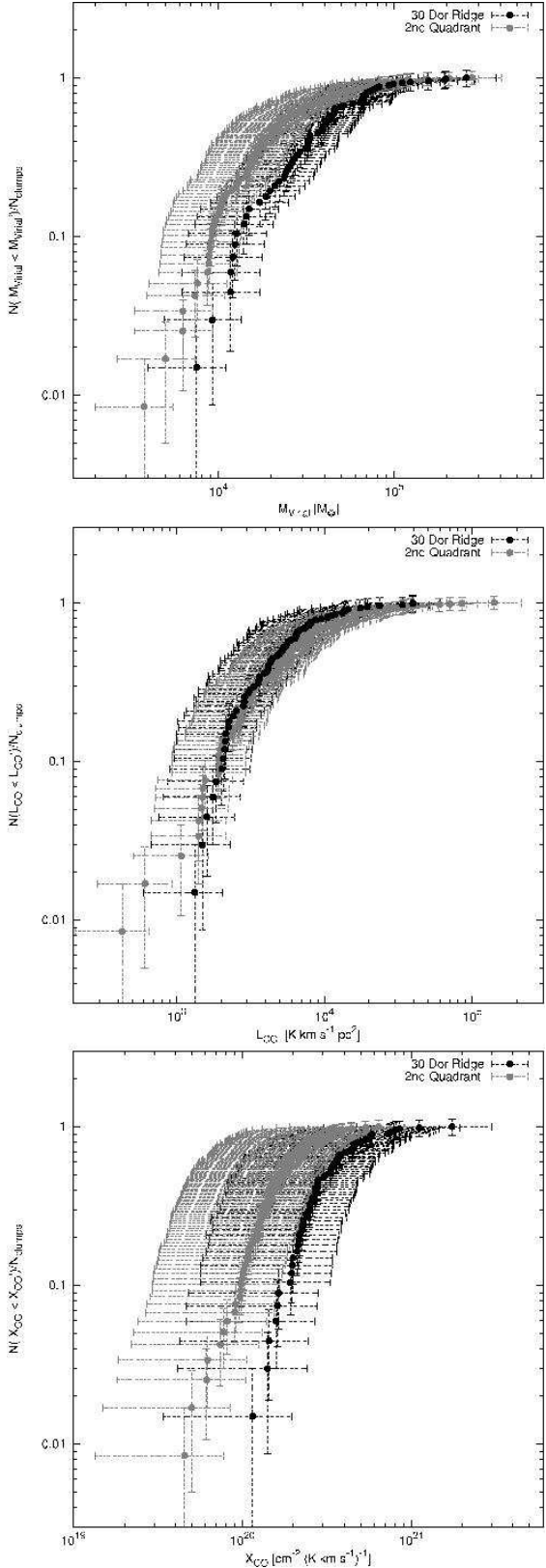}  
   \caption{Cumulative histograms of the virial mass ($M_\mathrm{vir}$), the CO
       luminosity ($L_\mathrm{CO}$), and the CO-to-H$_2$ conversion
       factor ($X_\mathrm{CO}$) derived in the 30\,Dor Ridge and 2nd
       Quadrant. For all curves, the horizontal error bars reflect
       the uncertainties in the observed properties and the vertical
       error bars represent $\sqrt N$ counting errors.}\label{fig:hist_advanced}
\end{figure}

We compare the properties of clumps in the 30\,Dor Ridge with those of
the 2nd Quadrant Galactic data.  The mean values and standard
deviation of these properties are listed in Table~2 and their
cumulative histograms are shown in Figures~\ref{fig:hist_basic} and
\ref{fig:hist_advanced}.  Although measures of their dispersion and
central tendency are comparable, Kolmogorov-Smirnov (KS) tests between
the distributions of $\Delta v'$, $R$, and $T_{\rm mb}$ for the two
datasets indicate that the probabilities (P) that the cloud samples
are drawn from the same parent distribution are 0, 0.07, and 0,
respectively.  These low probabilities suggest that there are
differences between the distribution of CO clump properties in both
regions.  KS tests for the distributions of $L_\mathrm{CO}$,
$M_\mathrm{vir}$, and $X_\mathrm{CO}$ also result in low probabilities
(P=0.3, 0, and 0), mainly due to the dissimilarity between the
distribution of $\Delta v'$, $T_{\rm mb}$, and $R$ in the 30\,Dor
ridge and the 2nd Quadrant.  In Figures~\ref{fig:hist_basic} we can
see that clumps in the 30\,Dor Ridge have lower brightness
temperatures compared with clumps in the 2nd Quadrant. This might be
an indication that the CO beam filling factor is lower in the 30\,Dor
Ridge, which can be a result of its reduced metallicity. The 30\,Dor
Ridge clumps have larger FWHM linewidths: 50\% of clouds in the 2nd
Quadrant have FWHM linewidths less than 4 km s$^{-1}$ compared to only
30\% in the 30\,Dor Ridge. This is in agreement with the predictions
from the theory of \citet{McKee1989} (see Section~3.4.1).  Finally, the
30\,Dor Ridge also shows a deficit of clumps with $R < 8\,$pc compared
to the 2nd Quadrant. This might be the effect of CO photodissociation
that destroy or prevent the formation of small clumps. The differences
in $\Delta v'$, $T_{\rm mb}$, and $R$ result in lower CO luminosities
and larger virial masses in the 30\,Dor Ridge.  We caution that in
case of $\Delta v'$ and $R$ the observed variations are comparable to
the errors in the determinations of these quantities.

The mean value of the $X_\mathrm{CO}$ conversion factor in the 30\,Dor
molecular ridge is 3.9$\pm$2.5$\times10^{20}$ cm$^{-2}$ (K~km
s$^{-1})^{-1}$). This is almost twice the average value for
$X_\mathrm{CO}$ in the 2nd Quadrant ([2.0$\pm$1.0]$\times10^{20}$
(cm$^{-2}$ K~km s$^{-1})^{-1}$), although we note that the dispersion
of $X_\mathrm{CO}$ values for the clumps in both datasets is quite
large.  Table~2 and Figure~\ref{fig:hist_advanced} show that the
average virial mass of clumps in the molecular ridge is $\sim 45$\%
larger than the average virial mass of the 2nd Quadrant clouds, but
that the average CO luminosity is $\sim 45$\% lower for the molecular
ridge clumps. This differences account for the factor of $\sim$2
larger value of $X_\mathrm{CO}$ in the 30\,Dor molecular ridge.  The
average value of $X_\mathrm{CO}$ in the 30\,Dor molecular ridge is
consistent with the 4.3$\times$10$^{20}$ cm$^{-2}$ (K~km
s$^{-1})^{-1}$ derived by \citet{Israel2003} using SEST observations
of several molecular cloud associated with H\,{\sc ii} regions in the
LMC (not including clouds in the 30\,Dor Ridge).

\subsubsection{Empirical Relations}

Several studies of molecular clouds in our Galaxy have shown that
there are empirical scaling relations between the basic observable
physical properties of molecular clouds
\citep[e.g.][]{Larson81,Myers83,Dame86,Solomon87}.  These relations,
often referred to as ``Larson's Laws'', link the FWHM linewidth
and size of a cloud, its CO luminosity and FWHM linewidth, its virial
mass and CO luminosity, and its CO luminosity and radius (note that
only two of these relations are independent). \citet{Solomon87}, for
example, found that clouds in the inner Milky Way follow power-laws of
the form $\Delta v$ $\propto R^{0.5}$, $L_\mathrm{CO} \propto
\Delta v^{5}$, $M_\mathrm{vir}$ $\propto$ $L_\mathrm{CO}^{0.81}$, and
$L_\mathrm{CO} \propto R^{2.5}$.  Power-laws with similar slopes have
already been observed for LMC molecular clouds, with some evidence
that clouds in the LMC have lower CO luminosities compared to Galactic
clouds with a similar FHWM linewidth
\citep[e.g.][]{Cohen88,Garay93,Johansson98,Mizuno01}. This has been
previously interpreted as evidence that CO is significantly
photo-dissociated in the LMC.

\begin{deluxetable}{lccc}
\label{sec:comp-with-galact}
\tablecaption{Mean Values of Derived CO Clump Parameters}
\tablehead{\colhead{Parameter} & \colhead{Mean} & \colhead{STDV} & \colhead{Units}} 
\startdata
30\,Dor Ridge\\
\hline
$X_\mathrm{CO}$&  3.9 & 2.5 & 10$^{20}$ cm$^{-2}$ (K~km s$^{-1})^{-1}$ \\ 
$M_\mathrm{vir}$& 5.2 & 4.7 & $10^{4}$ M$_\sun$ \\ 
$L_\mathrm{CO}$&  7.6 & 8.3 & $10^{3}$ K~km s$^{-1}$ pc$^{2}$  \\ 
$R$& 10 & 3 & pc\\ 
$\Delta v'$  & 4.6 & 1.8 & km s$^{-1}$ \\ 
$T_{\rm mb}$ &1.7 & 0.9 & K  \\

\hline
2nd Quadrant\\
\hline
$X_\mathrm{CO}$& 2.0 & 1.0 & 10$^{20}$ cm$^{-2}$ (K~km s$^{-1})^{-1}$ \\ 
$M_\mathrm{vir}$& 3.6 & 3.9 & $10^{4}$ M$_\sun$ \\  
$L_\mathrm{CO}$& 11 & 18  & $10^{3}$ K~km s$^{-1}$ pc$^{2}$\\ 
$R$& 10  & 3.8 & pc \\ 
$\Delta v'$& 3.8 & 1.4  & km s$^{-1}$ \\
$T_{\rm mb}$ &  3.0 & 1.6 & K
\enddata
\end{deluxetable}

\begin{figure*}[t]
       \centering \includegraphics[width=1.25\textwidth,angle=0]{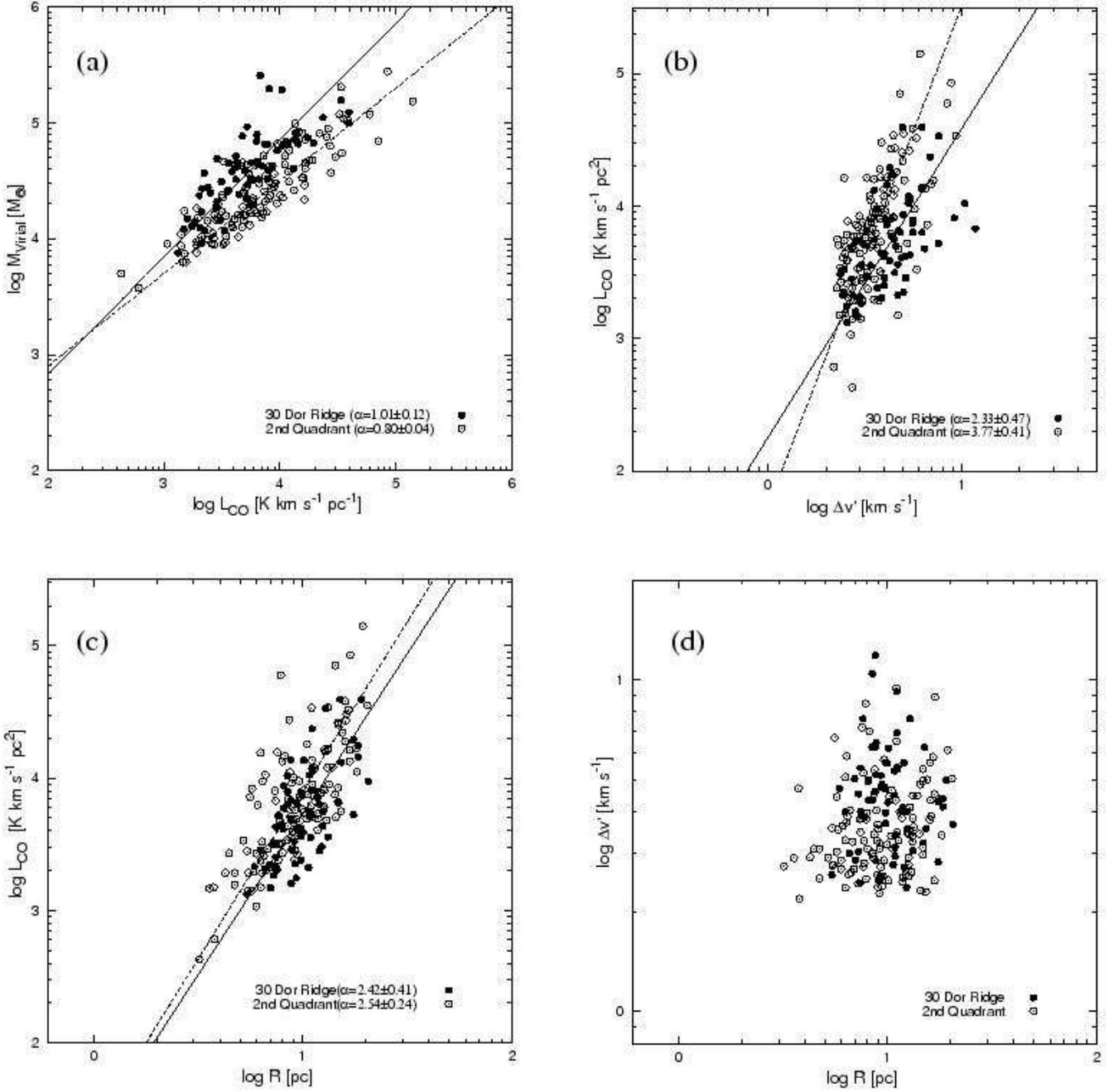}
     \caption{ The virial mass-CO luminosity, CO luminosity-FWHM linewidth, CO luminosity-clump size, and FWHM linewidth-size
relations for the identified clumps in the 30\,Dor molecular
ridge. For comparison, we also include the same relations for the
clumps identified in the 2nd Quadrant.  The thick and thin lines
correspond to fits to the relations for the 30\,Dor ridge and 2nd
Quadrant, respectively. For clarity, we do not show error bars. The typical errors in virial mass, CO luminosity, FWHM linewith, and size are 50\%, 50\%, 30\%, and 20\% respectively. }
\label{fig:relations}
\end{figure*}

Figure\,\ref{fig:relations} shows the observed relations between size,
velocity dispersion, CO luminosity and virial mass for the clumps
identified in the 30\,Dor molecular ridge and in the 2nd Quadrant.
The virial mass-CO luminosity relation is shown in
Figure\,\ref{fig:relations}a. The orthogonal BES fit\footnote{We fit
  power-laws using the orthogonal bi-variate error and intrinsic
  scatter method (BES; \citealt{Akritas1996}). Available at {\tt
  www.astro.wisc.edu/$\sim$mab/archive/stats/stats.html}.  } 
 results in

\begin{equation}
M_\mathrm{vir} =
10^{(0.8\pm0.4)}L_\mathrm{CO}^{(1.0\pm0.1)}
\end{equation}
 for the 30\,Dor molecular ridge, and 
\begin{equation}
M_\mathrm{vir} =
10^{(1.3\pm0.2)}L_\mathrm{CO}^{(0.8\pm0.04)} 
\end{equation}
for the 2nd Quadrant.  The relation is slightly steeper for clumps in
  the 30\,Dor molecular ridge compared to the molecular clouds in the
  2nd Quadrant, such that the larger ($M>5\times10^4$\,M$_{\odot}$)
  clumps in the 30\,Dor molecular ridge appear to be underluminous by
  a factor of $\sim$2.4, relative to a 2nd Quadrant cloud with a
  similar virial mass.  In Section~\ref{sec:clump-mass-spectrum} we
  will see that the number of massive clumps is larger in regions with
  stronger FUV radiation field ($\chi > 10$) than in regions with
  weaker FUV radiation field ($\chi < 10$).  This suggests that
  effects of CO photodissociation (enhanced by the FUV radiation
  field) might be more significant in high-mass rather than in
  low-mass clumps.  This might explain the apparent underluminosity of
  clumps with large virial masses in the 30\,Dor Ridge.  We note that
  the best-fitting virial mass-CO luminosity relation for clumps in
  the 30\,Dor molecular ridge is linear, which implies that the value
  of $X_{\rm CO}$ ($\propto M_\mathrm{vir}/L_\mathrm{CO}$) is
  independent of the clump mass. In the 2nd Quadrant, by contrast, the
  value of $X_{\rm CO}$ decreases with increasing mass. The variation
  in $X_{\rm CO}$ is quite small, however, corresponding to a factor
  of 0.3 decrease in $X_{\rm CO}$ over roughly two orders of magnitude
  in mass.

Figure\,\ref{fig:relations}b shows the FWHM linewidth-CO luminosity
relation. The fits are
\begin{equation}
L_\mathrm{CO}=10^{(2.2\pm0.3)}\Delta v'^{(2.3\pm0.5)} 
\end{equation}
for
the 30\,Dor ridge, and 
\begin{equation}
L_\mathrm{CO} =
10^{(1.7\pm0.2)}\Delta  v'^{(3.8\pm0.4)} 
\end{equation}
for the 2nd Quadrant.  The 30\,Dor ridge shows a shallower slope
compared with that for  outer Galaxy clouds.  Clumps in the
30\,Dor ridge with large FWHM linewidths ($\Delta v' > 3$\,km
s$^{-1}$) have CO luminosities that are a factor of 1.4 lower than
those of clumps in the 2nd Quadrant with similar $\Delta  v'$.

The CO luminosity-size relation is shown in
Figure\,\ref{fig:relations}c.  The fits are
\begin{equation}
L_\mathrm{CO} = 10^{(1.3\pm0.4)}R^{(2.4\pm0.4)} 
\end{equation}
for the
30\,Dor ridge, and 
\begin{equation}
L_\mathrm{CO} = 10^{(1.4\pm0.2)}
R^{(2.5\pm0.2)} 
\end{equation}
for the 2nd Quadrant. Both 30\,Dor Ridge and the 2nd Quadrant show a
similar slope. This slope is close to $L_\mathrm{CO} \propto R^{2.5}$,
which is expected for clouds in virial equilibrium that follow
$M_\mathrm{vir}$ $\propto$ $L_\mathrm{CO}^{0.8}$ and $\Delta v'
\propto R^{0.5}$. The best-fitting CO luminosity-size relations that
we determine imply that the mass surface density of the clumps
increases slowly with increasing clump mass, although the average mass
surface density for the large clumps in both datasets is in good
agreement with the canonical value for molecular clouds in the inner
Milky Way \citep[170~$M_{\odot}$~pc$^{-2}$, assuming the average
$X_{\rm CO}$ values that were estimated above][but see
\citet{Heyer2008} for evidence that this should be revised downwards
to $\sim 100~M_{\odot}$~pc$^{-2}$]{Solomon87}.

Unfortunately, we are not able to fit a power-law to the FWHM
linewidth-size relation shown in Figure\,\ref{fig:relations}d, as the
dynamic range in $\Delta v'$ and $R$ is insufficient. Nevertheless,
the close agreement between the mean $X_\mathrm{CO}$ derived from our
analysis of the CO emission and the value determined by
\citet{Dobashi2008} from dust extinction measurements suggests that
the clumps in the 30\,Dor molecular ridge are indeed close to virial
equilibrium.  As the clump properties do not show any systematic
variation along the ridge (Figure~\ref{fig:xco}), we conclude that the
degree of virialization is uniform for all clumps \citep[see
  also][]{Indebetouw2008}. 
Although, we see in Figure\,\ref{fig:relations}d that the clumps in
the 30\,Dor ridge and the 2nd Quadrant occupy the same region in the
parameter space, Table\,2 and Figure~\ref{fig:hist_basic} suggests
that clumps in the 30\,Dor Ridge have slightly larger $\Delta v'$
relative to the 2nd Quadrant. This is in apparent agreement with the
theory of photoionization-regulated star formation proposed by
\citet{McKee1989}, which predicts that molecular clouds in low
metallicity galaxies will have larger velocity dispersions than clouds
of a similar size with solar metallicity. Nonetheless, we note that
this prediction assumes that the filling fraction of the CO emission,
the densities of the CO-emitting clumps and the clump-to-cloud
extinction ratios are similar for molecular clouds in different
galaxies, which may not be the case \citep[e.g.][]{Heyer2008}.

   In conclusion, we find small differences between the scaling
  relations exhibited by the clump properties in the outer Galaxy and
  in the 30\,Dor molecular ridge.  This suggests that the basic
  physical properties of the molecular clumps are not especially
  sensitive to variations in the metallicity of the interstellar gas,
  or that the difference in metallicity between the 30\,Dor molecular
  ridge and clouds in the outer Galaxy is too small to produce
  significant disparities between the clump properties. A similar
  result has been obtained across a wider range of metallicities by
  \citet{Bolatto2008} (see also \citealt{Leroy2006} and
  \citealt{Sheth2008}).

\subsection{Clump-Mass Spectrum}
\label{sec:clump-mass-spectrum}

A general property of molecular clouds is that they show self-similar
structure over a wide range of scales.  This hierarchical, clumpy
structure may be quantified through the clump-mass spectrum of
molecular clouds, which describes how the molecular gas is distributed
across low- and high-mass substructures. Velocity-channel maps of
molecular clouds in the Milky Way
\citep[e.g.][]{Williams94,ElmegreenFalgalore96,Stutzki88,Kramer98,Heithausen98,Heyer98}
and in external galaxies
\citep{Staminirivic2000,Staminirovic2001,Fukui01,Kim2007} can be
decomposed into clumps that exhibit a power-law mass spectrum of the
form $dN/dM \propto M^{\alpha}$, where the exponent, $\alpha$, is
typically observed to lie between $-1.4$ to $-1.9$. This value of
$\alpha$, which indicates that the molecular gas is preferentially
distributed within massive substructures, holds over several orders of
magnitude of clump mass \citep[e.g.][]{Heithausen98}, and is
independent of the star-forming nature of the clouds
\citep[e.g.][]{Williams94}.

\begin{figure}
          \centering \includegraphics[width=0.6\textwidth,angle=0]{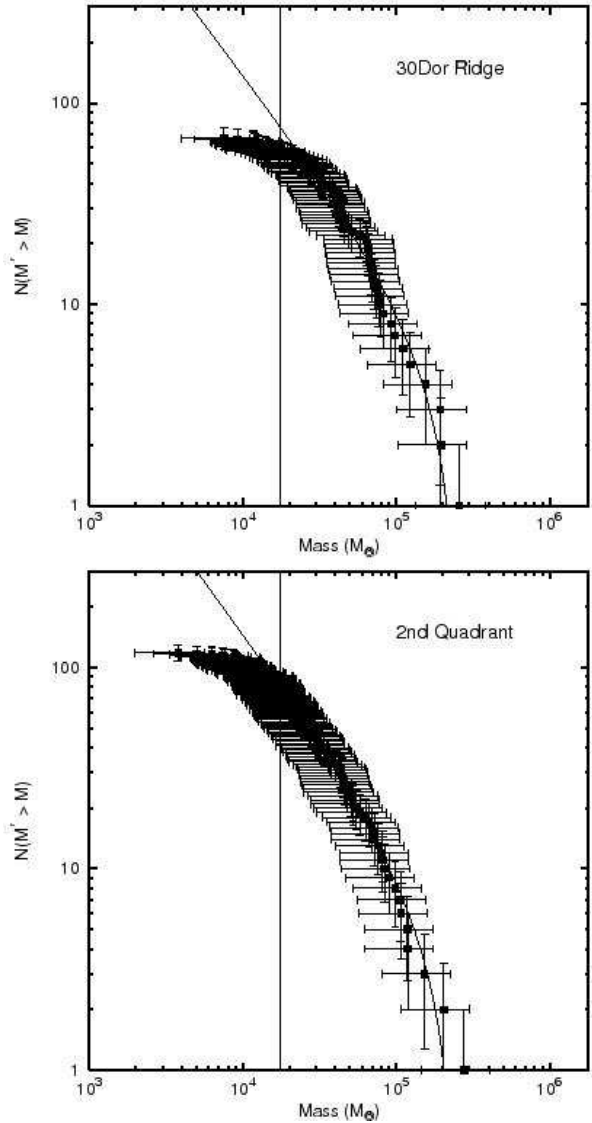}
     \caption{Cumulative clump-mass spectra for the ({\it upper
       panel}) 30\,Dor ridge and ({\it lower panel}) the 2nd
       Quadrant. The lines over the
       clump spectra represents the best truncated power-law fit over
       a range of masses above the completeness limit ($M > 1.76
       \times 10^{4}$M$_\odot$; vertical line). }
\label{fig:clum_mass_spectrum}
\end{figure}

In this Section, we study whether the metallicity and FUV radiation
field modify the distribution of clump masses in the ISM of the
LMC. For this analysis, we calculate the clump-mass spectrum of the
30\,Dor molecular ridge. We also calculate the mass spectrum of the
clumps identified in the 2nd Quadrant, in order to compare the mass
spectrum of clumps in low-metallicity and solar-metallicity gas,
avoiding systematic effects produced by the selection of a particular
decomposition method.

In the literature, the clump-mass spectrum based on $^{12}$CO data is
often calculated using clump masses derived from the clump luminosity
by applying a uniform $X_\mathrm{CO}$ conversion factor.  Since the
scatter in $X_\mathrm{CO}$ is large
(Section~\ref{sec:molec-cloud-prop}), in this study we prefer to use
virial masses for the calculation of the mass spectrum.  The total
virial mass of the 30\,Dor molecular ridge is about 3.5$\times10^{6}$\,M$_\odot$.

We express the clump-mass spectrum in its cumulative form rather than
in its differential form. This is motivated by the relatively low
number of identified clumps ($N = 67$) in the 30\,Dor ridge.  For such
a low number of clumps, the slope of the differential clump-mass
spectrum is sensitive to the binning process (which is not required by
the cumulative mass spectrum).  Following \citet{Rosolowsky2005}, we
fit a truncated power law of the form

\begin{equation}
\label{eq:3}
N(M'>M)=N_0 \left[ \left ( \frac{M}{M_0} \right )^{\alpha+1} -1 \right
],
\end{equation}

to our cumulative clump mass spectrum, where $M_0$ is the maximum mass
of the distribution and $N_0$ is the number of clumps with masses
larger than the mass where the distribution shows significant
deviation from a power-law, given by $2^{1/\alpha+1}M_0$.  A
truncation at high masses in the cumulative mass spectrum can either
be a real effect (i.e. due to a physical mechanism that regulates
cloud growth) or the result of finite sampling
\citep[e.g.][]{Mckee1997}.

\begin{figure}
          \centering \includegraphics[width=0.6\textwidth,angle=0]{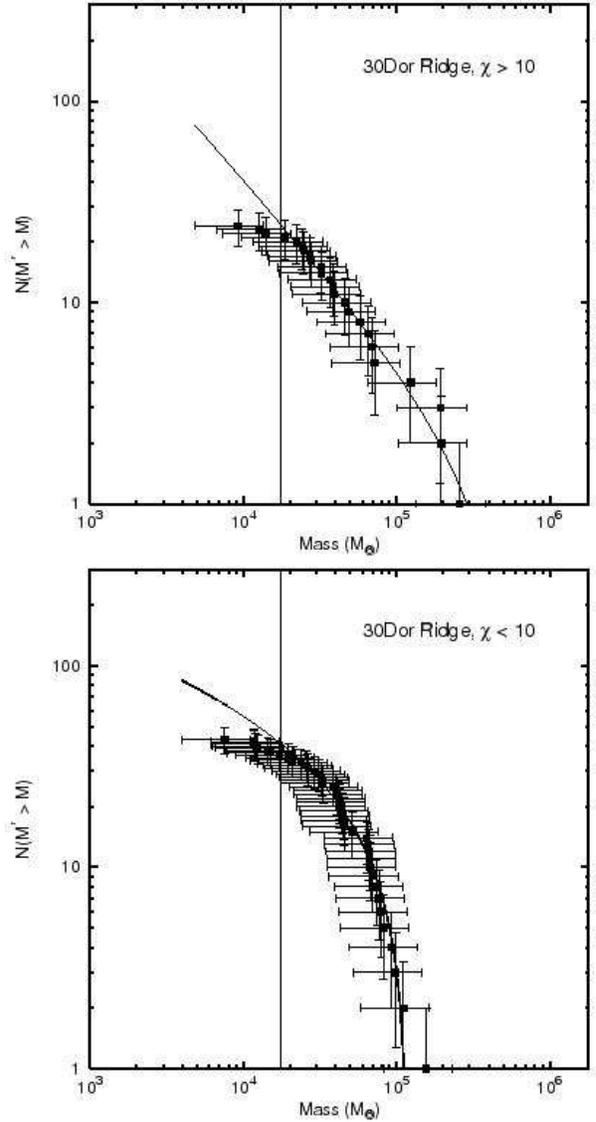}
     \caption{ Cumulative clump-mass spectra for the 30\,Dor ridge for
       clumps with $\chi > 10$ ({\it upper panel}) and with
       $\chi < 10$ ({\it lower panel}).  The lines over the
       clump spectra represents the best truncated power-law fit over
       a range of masses above the completeness limit ($M \gtrsim 2
       \times 10^{4}$M$_\odot$; vertical line).}
\label{fig:clumps_mass_vs_chi}
\end{figure}

\begin{deluxetable}{lccc}
\label{sec:clump-mass-spectrum-1}
\tablecaption{Parameters of Mass Spectra for clumps in the 30\,Dor ridge and
  2nd Quadrant}
\tablenum{3}
\tablehead{\colhead{Dataset} & \colhead{$\alpha$} & \colhead{$N_0$} 
& \colhead{$M_0/(10^5\,M_\sun)$}} 

\startdata

30\,Dor ridge &                      -2.0 $\pm$ 0.4   & 5.9 $\pm$ 5.9  & 2.5 $\pm$ 0.9\\ 
2nd Quadrant &                       -2.0 $\pm$ 0.3   & 5.8 $\pm$ 6.2  & 2.4 $\pm$ 1.1\\ 
30\,Dor ridge ($\chi>10$) &  -1.8 $\pm$ 0.3   & 1.3 $\pm$ 2.7  & 5.5 $\pm$ 1.6 \\ 
30\,Dor ridge ($\chi<10$) &  -1.2 $\pm$ 0.3   & 97 $\pm$ 10 & 1.2 $\pm$ 0.3 \\ 
\enddata
\end{deluxetable}

We use the error-in-variables procedure described by
\citet{Rosolowsky2005} to fit a power-law to cumulative mass
spectra. This approach has the advantage that it accounts for both the
$(\Delta N)^{1/2}$ statistical error in the distributions and the
errors in the measurements of the clump masses.

The introduction of a cutoff criterion that rejects noise features and
clouds with low CO emission in the decomposition implies that some
low-mass clumps are missing from the clump-mass spectrum. This effect
produces a turnover in the mass spectra at low masses.  We assume that
the incompleteness of our sample begins to be significant when the
clump mass is lower than 5 times the minimum mass that can be selected
in our sample based on the criteria described in
Section~\ref{sec:analysis} for the 30\,Dor ridge
($3.5\times10^{3}$\,M$_\odot$).  We therefore fit a truncated power
law to the cumulative clump mass distribution using clumps with masses
greater than our adopted completeness limit of $M^{c}_\mathrm{vir}
\simeq 2\times10^{4}$\,M$_\odot$. For consistency, we use this
completeness limit for both the 30\,Dor molecular ridge and the 2nd
Quadrant datasets.

Fig. \ref{fig:clum_mass_spectrum} shows the cumulative clump-mass
spectra of the 30\,Dor molecular ridge and the 2nd Quadrant. The
results of the fits are given in Table~3.

The exponents of the power-law fits to the clump-mass spectra of the
30\,Dor ridge and in the 2nd Quadrant are similar within the
errors. This suggests that the reduced metal abundance in the LMC does
not significantly affect the clump-mass spectrum of molecular clouds
across the observed range of clump masses.  The fits are also
consistent with the mass spectra that were derived by
\citet{Rosolowsky2005} for clouds in the outer Milky Way
($\alpha=-2.29\pm$0.08, $N_0$=4.5$\pm$3.5, and $M_0/10^{5}
M_\odot$=2.9$\pm$1.0; data from \citealt{Brunt2003}) and in the entire
LMC ($\alpha=-1.71\pm$0.19, $N_0$=10$\pm$6.5, and $M_0/10^{5}
M_\odot$=23$\pm$4.6; data from \citealt{Mizuno2001}) using the same
fitting procedure and virial estimates for the molecular cloud
masses. Note that the truncated power-law fit by
\citet{Rosolowsky2005} to the \citet{Mizuno2001} LMC cloud catalog was
based in virial mass measurements that do not account for beam
deconvolution. If we fit a truncated power law to our molecular ridge
data using undeconvolved measurements for the clump properties, we
obtain a slope for the mass spectrum of $\alpha=-2.2\pm0.3$, which is
still close to  their results.  Some of the scatter in the
derived slopes is probably due to the different decomposition
algorithms used to identify cloud structures in the various studies,
although we note that the cloud mass spectrum constructed from the LMC
cloud catalogue includes clouds across the LMC, not just the molecular
ridge region. It also extends to larger cloud masses and imposes a
higher completeness limit than we have adopted for our analysis.  
  Previous works that find flatter slopes for the clump mass spectrum
  in the outer Galaxy \citep[e.g. $\alpha=-1.45$ ][]{May1997} are not
  directly comparable to our results, as the derived slope is strongly
  dependent on how the mass spectrum is constructed
  \citep{Rosolowsky2005}.  For this reason, we prefer to compare our
  results with independent analyzes that have employed the same
  representation of the mass spectrum that we have used here.

 We verified that the use of virial masses instead of masses
  derived from CO luminosities and a constant value for the $X_{\rm
    CO}$ conversion factor does not have a significant impact on the
  clump mass spectra of the 30\,Dor molecular ridge and 2nd Quadrant.
  For clumps in the molecular ridge, assuming $X_{\rm
    CO}$=3.9$\times10^{20}$ cm$^{-2}$ (K~km s$^{-1})^{-1}$ (see
  Section~\ref{sec:molec-cloud-prop}), the best-fitting truncated
  power law to the clump mass spectrum yields $\alpha=-2.2\pm0.3$,
  $N_0=$1.5$\pm2.8$, and $M_0/10^5$=2.4$\pm$0.7. For the 2nd
  Quadrant, the best fit gives $\alpha=-1.9\pm0.2$,
  $N_0=$5$\pm8$, and $M_0/10^5$=4.8$\pm$2.7, assuming $X_{\rm
    CO}$=2.0$\times10^{20}$ cm$^{-2}$ (K~km s$^{-1})^{-1}$.  For both
  datasets, the fits are consistent with those determined for the
  cumulative mass spectra constructed from virial estimates of the
  clump masses.

\citet{Fukui01} obtain a slope for the molecular cloud mass spectrum
in the entire LMC of $-1.9$, covering a range of cloud masses from
$\sim$ $8\times10^{4}$ to $3\times10^{6}$\,M$_\odot$.  Our analysis
shows that the power-law relation extends to even lower masses
($\sim2\times10^{4}$\,M$_\odot$), suggesting that the ISM in the LMC
might show the self-similarity over a wide range of scales, as observed
for molecular clouds in the Milky Way \citep{Heithausen98}.

In order to study the influence of the strength of the FUV radiation
field on the clump mass spectrum, we divide the clumps in the 30\,Dor
molecular ridge into low-FUV ($\chi<10$) and high-FUV ($\chi>10$)
subsamples. The resulting clump-mass spectra are shown in
Figure~\ref{fig:clumps_mass_vs_chi} and the results of the truncated
power-law fits to the spectra are listed in Table~3.  The derived
slope for the $\chi<10$ subsample (-1.2$\pm$0.3) is dominated by
low mass clumps, and the mass spectrum steepens considerably for
larger clump masses. If we fit the spectrum with a simple power-law (i.e. without
truncation) of the form
\begin{equation}
N(M'>M)=\left ( \frac{M}{M_0} \right )^{\alpha+1},
\end{equation}
where $M_0$ and $\alpha$ represent the same quantities used in
Equation~(\ref{eq:3}), we obtain a much steeper slope of
$\alpha=-2.7\pm0.2$.  Figure~\ref{fig:clumps_mass_vs_chi} shows that
there is a deficit of high-mass clumps with $\chi<10$.  This might be
because massive stars form preferentially in massive rather than in
low-mass clumps. Since the strength of the local FUV field depends on
embedded massive stars, the average FUV field will be larger for
massive clumps than for low-mass clumps, leading to the apparent
deficit of large clumps with $\chi<10$.

\section{Discussion}
\label{sec:discussion}

As outlined in the previous sections, the properties of molecular
clouds in the 30\,Dor molecular ridge, as traced by their CO emission,
are not significantly affected by the varying intensity of the FUV
radiation field interacting with the reduced-metallicity gas of the
LMC. The requisite physical conditions for CO emission may provide a
possible explanation for this lack of dependence. As the
photodissociation of CO is enhanced in low-metallicity systems, CO
emission is only present in regions with large H$_2$ column densities.
Thus, CO observations might only trace regions that are well-shielded
from the FUV radiation, as suggested by e.g. \citet{Johansson98}.  In
this case, the 30\,Dor molecular ridge might contain a substantial
fraction of H$_2$ coexisting with other carbon species such as C and
C$^{+}$, and this ``CO-dark'' fraction of the molecular gas may be
larger than for clouds in the Milky Way \citep[10 to
50\%]{Grenier2005}. Indeed, \citet{Pineda2008} recently found that the
column density of CO toward LMC N159W is very much reduced compared to
the column densities of C and C+.  This might be a reason why studies
of molecular clouds in low-metallicity galaxies based on dust emission
(which should trace the entire molecular content of clouds once the
contribution from H\,{\sc i} is subtracted) usually measure larger
molecular masses than studies based on CO emission
\citep[e.g.][]{Israel97,Rubio2004,Leroy2007,Bot2007}. However, the
measurement of molecular gas masses from dust emission usually assumes
that the dust and gas are well mixed, and further relies on estimates
of the dust-to-gas ratio, dust temperature, and optical properties of
dust grains.  Using near infrared extinction, a method that is
independent of the cloud's dynamical state and of cloud decomposition
method, \citet{Dobashi2008} obtained an average $X_\mathrm{CO}$
conversion factor of 2.26$\times$10$^{20}$ cm$^{-2}$ (K~km
s$^{-1})^{-1}$ in the entire LMC which is a factor of $\sim$1.7 lower than
our mean value of the 30\,Dor ridge\footnote{The value of $X_{\rm CO}$
derived from dust extinction depends on the adopted extinction to
total column density ratio. When using the values adopted by
\citet{Imara2007}, their $X_{\rm CO}$ factor is 3.55$\times$10$^{20}$
cm$^{-2}$ (K~km s$^{-1})^{-1}$, in closer agreement with our
determination.}.  Note that the \citet{Dobashi2008} value refers to an
H$_2$ column density associated with CO-emitting gas.

\begin{figure}
  \centering
  \includegraphics[width=0.48\textwidth,angle=0]{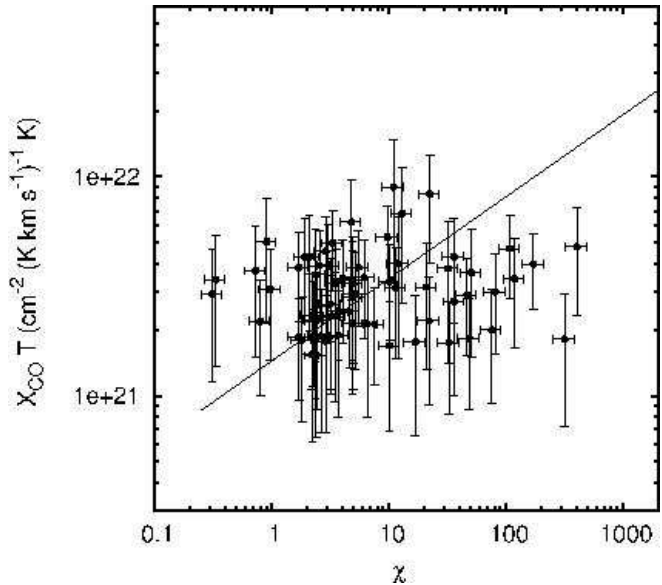}
     \caption{Relation between $X_\mathrm{CO}$, $T$ and
       the strength of the FUV radiation field for the clumps in the
       30\,Dor ridge dataset (solid black boxes). The straight line
       is the relation predicted by \citet{Elmegreen89}.  }
\label{fig:elmegreen}
\end{figure}

\citet{Elmegreen89} modelled the relation between $X_\mathrm{CO}$ and
the strength of the FUV field as $X_\mathrm{CO} \propto \chi^{3/8}/T$,
where $T$ is the $^{12}$CO $J = 1\to0$ brightness temperature.  We can test this
prediction using our data (see Figure~\ref{fig:elmegreen}).  Note that
the relation presented by \citet{Elmegreen89} also depends weakly on
the clump mass and on the ambient pressure, although we ignore their
contribution here.  We corrected the observed $^{12}$CO $J = 1\to0$ peak
brightness temperature assuming a beam filling factor of 1/6
\citep{Pineda2008}.  Figure~\ref{fig:elmegreen} shows that our data is
inconsistent ($\chi^2\simeq$254) with the theoretical prediction by
\citet{Elmegreen89}. Our result that $X_\mathrm{CO}$ is not correlated
with $\chi$ for the molecular clumps in our survey also appears to
contradict the $X_\mathrm{CO}$ $\propto \chi$ relation obtained by
\citet{Israel97}. However, the range of FUV radiation field estimates
in that paper (which are derived from infrared observations and
expressed in terms of the radiative energy per nucleon) is quite
limited, whereas our estimate of $\chi$ extends over about three
orders of magnitude.  We also note that the cloud mass estimates from
\citet{Israel97} are based on dust observations.  If $X_\mathrm{CO}$
is independent of $\chi$ for CO observations and $X_\mathrm{CO}
\propto \chi$ for dust observations, this would imply that the amount
of ``CO-dark'' molecular gas increases with the strength of the FUV
radiation field.  However, this would seem to be in contradiction with the
results of \citet{Dobashi2008}, who found no systematic variation of
the extinction-based $X_\mathrm{CO}$ as a function of distance to
30\,Doradus (i.e. from quiescent regions to regions exposed to strong
FUV radiation fields).

 The observed relation between $X_{\rm CO}$ and $\chi$ can be
  reconciled with the theory of \citet{Elmegreen89} if the kinetic
  temperature scales with the FUV radiation field. Since the CO
  brightness temperature does not show significant variations with
  $\chi$ (Figure~\ref{fig:tmb_dv_r}), this would require that the beam
  filling decreases with $\chi$ throughout the 30\,Dor Ridge. It is
  plausible that the kinetic is larger in regions with higher $\chi$,
  since photoelectric heating is likely to be the dominant heating
  mechanism. For example, if clouds with $\chi=1$ have $T_{\rm
  kin}=10$\,K, kinetic temperatures of 32\,K and 66\,K are required
  for N159W ($\chi=173$) and 30\,Dor-10 ($\chi=407$), respectively, to
  compensate any increase of $X_{\rm CO}$ with $\chi$. These kinetic
  temperatures are consistent with estimations in the literature
  \citep[e.g.][]{Johansson98,Minamidani2007,Pineda2008}.  A systematic
  decrease of the beam filling with $\chi$ might be the result of an
  enhanced rate of CO photodissociation which reduces the size of CO
  cores within the beam. However, other mechanism might also play a
  role e.g. cloud disruption by stellar winds from newly formed stars,
  etc.

\begin{figure}
  \centering
  \includegraphics[width=0.65\textwidth,angle=0]{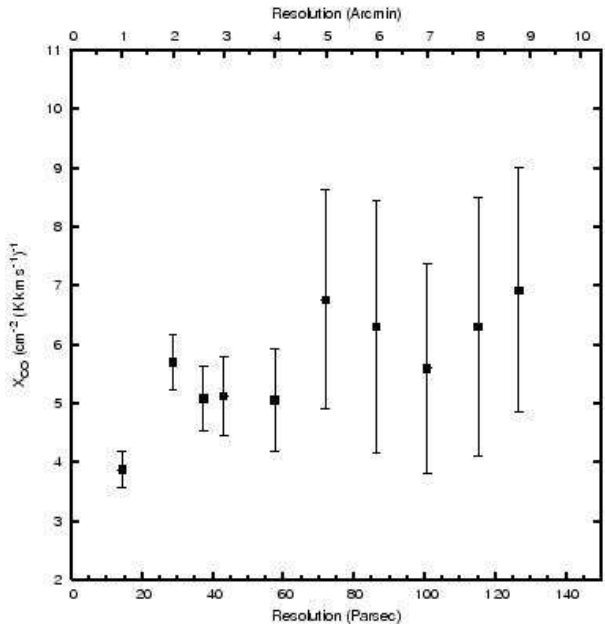}
     \caption{ Average $X_\mathrm{CO}$ obtained for the 30\,Dor
       ridge map for  different angular resolutions. The error bars
       correspond to the standard error of the mean.  }\label{fig:res}
\end{figure}

The average $X_\mathrm{CO}$ conversion factor found in our survey of
the 30\,Dor molecular ridge is a factor of $\sim 2$ larger than that
found in  outer Galaxy clouds, a difference that is smaller
compared with previous studies of clouds in the LMC at coarser angular
resolutions.  As we can see in Fig~\ref{fig:ridge}, the molecular
ridge is composed of a few cloud complexes with linear scales between
$100-300$\,pc. These cloud complexes show substructure characterized
by compact clumps connected by low-surface brightness gas on scales
between $10-30$\,pc.  Since {\tt GAUSSCLUMPS} decomposes the CO
emission and identifies clumps by fitting local CO intensity peaks,
this substructure is the one characterized in the present paper.
Therefore, the average $X_\mathrm{CO}$ estimated in our survey refers
to scales between $10-30$\,pc.  The large cloud complexes do not
necessarily have the same $X_\mathrm{CO}$ as the individual clumps.
Although increases in the clump size are tracked by increasing
estimates for the virial mass, the CO luminosity will not show an
equivalent increase if a significant quantity of low-surface
brightness inter-clump gas is present. In this case, the appropriate
$X_\mathrm{CO}$ conversion factor will be higher for larger scale
molecular gas structures, as shown by e.g.  \citet{Bolatto03}.
Observations at coarser angular resolution trace these large-scale
complexes rather than small clumps, which helps to explain the larger
values of $X_\mathrm{CO}$ derived by previous surveys in the LMC.  An
example of how $X_\mathrm{CO}$ increases with the resolution of the
observations is shown in Figure~\ref{fig:res}, where we plot the
average $X_\mathrm{CO}$ obtained when {\tt GAUSSCLUMPS} is applied to
versions of the 30\,Dor molecular ridge datacube that have been
smoothed to different angular resolutions. We include the average
$X_\mathrm{CO}$ for 2.6\arcmin\,(40\,pc) and 8.8\arcmin\,(127\,pc),
corresponding to the resolution of the NANTEN \citep{Fukui2008} and
CfA 1.2m \citep{Cohen88} telescopes.  At the resolution of the NANTEN
survey, we obtain an average value of $X_\mathrm{CO}
\sim$5.1$\times10^{20}$cm$^{-2}$ (K~km s$^{-1})^{-1}$, which is
somewhat lower than the value obtained by \citet{Fukui2008},
$X_\mathrm{CO}= 7\times10^{20}$cm$^{-2}$ (K~km s$^{-1})^{-1}$. Our
result is in better agreement with the determination by
\citet{Blitz2007} for the same dataset, $X_\mathrm{CO} =
5.4\times10^{20}$cm$^{-2}$ (K~km s$^{-1})^{-1}$: the discrepancy
between the \citet{Blitz2007} and \citet{Fukui2008} may arise from a
different choice of input parameters for the cloud decomposition
algorithm ({\tt CLOUDPROPS;} \citealt[][]{Rosolowsky2006}). At the
resolution of the \citet{Cohen88} observations, we obtain
$X_\mathrm{CO}\simeq$6.9$\times10^{20}$cm$^{-2}$ (K~km s$^{-1})^{-1}$,
a factor of $\sim$1.6 lower than the value quoted by \citet{Cohen88}
($X_\mathrm{CO}\simeq11\times10^{20}$cm$^{-2}$ (K~km s$^{-1})^{-1}$,
assuming 1.8$\times10^{20}$cm$^{-2}$ (K~km s$^{-1})^{-1}$ for Milky
Way clouds). In summary, Figure~\ref{fig:res} suggests that scale of
the observed structures must be considered when the total molecular
mass of galaxies is calculated and that comparisons between
extragalactic and Galactic molecular cloud populations should adopt
the same decomposition and cloud identification methods.

\section{Summary and Conclusions}
\label{sec:conclusions}

We have presented a large-scale $^{12}\mathrm{CO}~J~= 1\to 0$ survey
of the 30 Dor molecular ridge in the Large Magellanic Cloud observed
with the Mopra telescope. From the analysis of these observations we
conclude the following:

\begin{enumerate}

\item We estimate the strength of the far-ultraviolet radiation field
  for the entire 30\,Dor molecular ridge, in order to study the
  dependence of CO cloud properties on this parameter. We find that CO
  clump properties are insensitive to changes in the strength of the
  FUV radiation field over almost 3 orders of magnitude in $\chi$.
    The observed relation between CO-to-H$_2$ conversion factor and
    the strength of the FUV field can be reconciled with the expected
    increase predicted by theory if the kinetic temperature also
    increases in regions with strong FUV fields. Since the CO
    brightness temperatures do not show any significant variation
    along the 30\,Dor molecular ridge, this would require that the
    angular filling factor of the CO emission shows a corresponding
    decrease with $\chi$.

\item Applying the same cloud decomposition method to clouds in the
  30\,Dor molecular ridge and in the outer Galaxy clouds in the 2nd
  Quadrant, the average CO-to-H$_2$ conversion factor obtained in our
  survey is $X_\mathrm{CO} \simeq 2$\,$X_\mathrm{MW}$.  This
  discrepancy is smaller than has been reported by previous studies of
  the CO emission in the LMC at lower angular resolution.

\item Studying the scaling relations between the physical properties
  of the observed clumps, we find that the molecular clouds in the
  30\,Dor ridge exhibit properties that are similar, but not
  identical, to those in the outer Galaxy. In particular, we find some
  evidence that clumps in the 30\,Dor molecular ridge have larger
  velocity dispersions and lower brightness temperatures than their
  outer Galaxy counterparts. Additonally, we find a deficit of small
  clumps in the 30\,Dor molecular ridge compared to the outer
  galaxy. Further studies of the emission from atomic hydrogen and
  other carbon species in the interclump gas within LMC molecular
  clouds would be useful for determining the physical origin of these
  slight variations. The degree of virialization of the clumps does
  not appear to show any significant variation along the molecular
  ridge.

\item An analysis of the clump mass spectra yields a power-law index
  of $\alpha = -2.0\pm0.4$ for the entire 30\,Dor molecular ridge. The
  clump mass spectrum is similar to the mass spectra obtained for
  Galactic clouds and for the entire LMC cloud catalog. Comparing
  different sub-regions with different average strengths of the FUV
  field, we find that the clump-mass spectrum is shallower in regions
  that are exposed to strong FUV fields. We propose that this is
  caused by the preferential formation of massive stars within large
  molecular clumps, rather than an effect of CO photo-dissociation in
  the low-metallicity molecular gas.

  \end{enumerate}

\acknowledgments

This work was supported by the Deutsche Forschungs Gemeinschaft (DFG)
via Grant SFB 494. We would like to thank Frank Israel for providing
the SEST data set used to check the calibration of our data, Thomas
Dame for providing the 2nd Quadrant dataset, and Erik Rosolowsky, for
helping us with {\tt CLOUDPROPS}, used to check our {\tt GAUSSCLUMPS}
decomposition, and with the error-in-variables procedure.  The
National Radio Astronomy Observatory is a facility of the National
Science Foundation operated under cooperative agreement by Associated
Universities, Inc.  We made use of the NASA/IPAC/IRAS/HiRES data
reduction facilities.  This research has made use of NASA's
Astrophysics Data System Abstract Service.

\bibliography{ms}

\begin{thebibliography}{88}
\expandafter\ifx\csname natexlab\endcsname\relax\def\natexlab#1{#1}\fi

\bibitem[{{Akritas} \& {Bershady}(1996)}]{Akritas1996}
{Akritas}, M.~G. \& {Bershady}, M.~A. 1996, \apj, 470, 706

\bibitem[{{Arimoto} {et~al.}(1996){Arimoto}, {Sofue}, \&
  {Tsujimoto}}]{Arimoto96}
{Arimoto}, N., {Sofue}, Y., \& {Tsujimoto}, T. 1996, \pasj, 48, 275

\bibitem[{{Aumann} {et~al.}(1990){Aumann}, {Fowler}, \& {Melnyk}}]{Aumann1990}
{Aumann}, H.~H., {Fowler}, J.~W., \& {Melnyk}, M. 1990, \aj, 99, 1674

\bibitem[{{Bakes} \& {Tielens}(1994)}]{BakesTielens94}
{Bakes}, E.~L.~O. \& {Tielens}, A.~G.~G.~M. 1994, \apj, 427, 822

\bibitem[{{Blitz} {et~al.}(2007){Blitz}, {Fukui}, {Kawamura}, {Leroy},
  {Mizuno}, \& {Rosolowsky}}]{Blitz2007}
{Blitz}, L., {Fukui}, Y., {Kawamura}, A., {Leroy}, A., {Mizuno}, N., \&
  {Rosolowsky}, E. 2007, in Protostars and Planets V, ed. B.~{Reipurth},
  D.~{Jewitt}, \& K.~{Keil}, 81--96

\bibitem[{{Bolatto} {et~al.}(2003){Bolatto}, {Leroy}, {Israel}, \&
  {Jackson}}]{Bolatto03}
{Bolatto}, A.~D., {Leroy}, A., {Israel}, F.~P., \& {Jackson}, J.~M. 2003, \apj,
  595, 167

\bibitem[{{Bolatto} {et~al.}(2008){Bolatto}, {Leroy}, {Rosolowsky}, {Walter},
  \& {Blitz}}]{Bolatto2008}
{Bolatto}, A.~D., {Leroy}, A.~K., {Rosolowsky}, E., {Walter}, F., \& {Blitz},
  L. 2008, \apj, 686, 948

\bibitem[{{Bot} {et~al.}(2007){Bot}, {Boulanger}, {Rubio}, \&
  {Rantakyro}}]{Bot2007}
{Bot}, C., {Boulanger}, F., {Rubio}, M., \& {Rantakyro}, F. 2007, \aap, 471,
  103

\bibitem[{{Brunt} {et~al.}(2003){Brunt}, {Kerton}, \& {Pomerleau}}]{Brunt2003}
{Brunt}, C.~M., {Kerton}, C.~R., \& {Pomerleau}, C. 2003, \apjs, 144, 47

\bibitem[{{Cohen} {et~al.}(1988){Cohen}, {Dame}, {Garay}, {Montani}, {Rubio},
  \& {Thaddeus}}]{Cohen88}
{Cohen}, R.~S., {Dame}, T.~M., {Garay}, G., {Montani}, J., {Rubio}, M., \&
  {Thaddeus}, P. 1988, \apjl, 331, L95

\bibitem[{{Dale} {et~al.}(2001){Dale}, {Helou}, {Contursi}, {Silbermann}, \&
  {Kolhatkar}}]{Dale2001}
{Dale}, D.~A., {Helou}, G., {Contursi}, A., {Silbermann}, N.~A., \&
  {Kolhatkar}, S. 2001, \apj, 549, 215

\bibitem[{{Dame} {et~al.}(1986){Dame}, {Elmegreen}, {Cohen}, \&
  {Thaddeus}}]{Dame86}
{Dame}, T.~M., {Elmegreen}, B.~G., {Cohen}, R.~S., \& {Thaddeus}, P. 1986,
  \apj, 305, 892

\bibitem[{{Dame} {et~al.}(2001){Dame}, {Hartmann}, \& {Thaddeus}}]{Dame01}
{Dame}, T.~M., {Hartmann}, D., \& {Thaddeus}, P. 2001, \apj, 547, 792

\bibitem[{{Dettmar} \& {Heithausen}(1989)}]{Dettmar89}
{Dettmar}, R.-J. \& {Heithausen}, A. 1989, \apjl, 344, L61

\bibitem[{{Dickel} {et~al.}(2005){Dickel}, {McIntyre}, {Gruendl}, \&
  {Milne}}]{Dickel05}
{Dickel}, J.~R., {McIntyre}, V.~J., {Gruendl}, R.~A., \& {Milne}, D.~K. 2005,
  \aj, 129, 790

\bibitem[{{Dobashi} {et~al.}(2008){Dobashi}, {Bernard}, {Hughes}, {Paradis},
  {Reach}, \& {Kawamura}}]{Dobashi2008}
{Dobashi}, K., {Bernard}, J.-P., {Hughes}, A., {Paradis}, D., {Reach}, W.~T.,
  \& {Kawamura}, A. 2008, \aap, 484, 205

\bibitem[{{Draine}(1978)}]{Draine78}
{Draine}, B.~T. 1978, \apjs, 36, 595

\bibitem[{{Dufour}(1984)}]{Dufour1984}
{Dufour}, R.~J. 1984, in IAU Symposium, Vol. 108, Structure and Evolution of
  the Magellanic Clouds, ed. S.~{van den Bergh} \& K.~S.~D. {Boer}, 353--360

\bibitem[{{Elmegreen}(1989)}]{Elmegreen89}
{Elmegreen}, B.~G. 1989, \apj, 338, 178

\bibitem[{{Elmegreen} \& {Falgarone}(1996)}]{ElmegreenFalgalore96}
{Elmegreen}, B.~G. \& {Falgarone}, E. 1996, \apj, 471, 816

\bibitem[{{Feast}(1999)}]{Feast99}
{Feast}, M. 1999, \pasp, 111, 775

\bibitem[{{Fukui} {et~al.}(2008){Fukui}, {Kawamura}, {Minamidani}, {Mizuno},
  {Kanai}, {Mizuno}, {Onishi}, {Yonekura}, {Mizuno}, {Ogawa}, \&
  {Rubio}}]{Fukui2008}
{Fukui}, Y., {Kawamura}, A., {Minamidani}, T., {Mizuno}, Y., {Kanai}, Y.,
  {Mizuno}, N., {Onishi}, T., {Yonekura}, Y., {Mizuno}, A., {Ogawa}, H., \&
  {Rubio}, M. 2008, \apjs, 178, 56

\bibitem[{{Fukui} {et~al.}(2001){Fukui}, {Mizuno}, {Yamaguchi}, {Mizuno}, \&
  {Onishi}}]{Fukui01}
{Fukui}, Y., {Mizuno}, N., {Yamaguchi}, R., {Mizuno}, A., \& {Onishi}, T. 2001,
  \pasj, 53, L41

\bibitem[{{Fukui} {et~al.}(1999){Fukui}, {Mizuno}, {Yamaguchi}, {Mizuno},
  {Onishi}, {Ogawa}, {Yonekura}, {Kawamura}, {Tachihara}, {Xiao}, {Yamaguchi},
  {Hara}, {Hayakawa}, {Kato}, {Abe}, {Saito}, {Mano}, {Matsunaga}, {Mine},
  {Moriguchi}, {Aoyama}, {Asayama}, {Yoshikawa}, \& {Rubio}}]{Fukui99}
{Fukui}, Y., {Mizuno}, N., {Yamaguchi}, R., {Mizuno}, A., {Onishi}, T.,
  {Ogawa}, H., {Yonekura}, Y., {Kawamura}, A., {Tachihara}, K., {Xiao}, K.,
  {Yamaguchi}, N., {Hara}, A., {Hayakawa}, T., {Kato}, S., {Abe}, R., {Saito},
  H., {Mano}, S., {Matsunaga}, K., {Mine}, Y., {Moriguchi}, Y., {Aoyama}, H.,
  {Asayama}, S.-i., {Yoshikawa}, N., \& {Rubio}, M. 1999, \pasj, 51, 745

\bibitem[{{Garay} {et~al.}(1993){Garay}, {Rubio}, {Ramirez}, {Johansson}, \&
  {Thaddeus}}]{Garay93}
{Garay}, G., {Rubio}, M., {Ramirez}, S., {Johansson}, L.~E.~B., \& {Thaddeus},
  P. 1993, \aap, 274, 743

\bibitem[{{Gaustad} {et~al.}(2001){Gaustad}, {McCullough}, {Rosing}, \& {Van
  Buren}}]{Gaustad01}
{Gaustad}, J.~E., {McCullough}, P.~R., {Rosing}, W., \& {Van Buren}, D. 2001,
  \pasp, 113, 1326

\bibitem[{{Grenier} {et~al.}(2005){Grenier}, {Casandjian}, \&
  {Terrier}}]{Grenier2005}
{Grenier}, I.~A., {Casandjian}, J.-M., \& {Terrier}, R. 2005, Science, 307,
  1292

\bibitem[{{Habing}(1968)}]{Habing68}
{Habing}, H.~J. 1968, \bain, 19, 421

\bibitem[{{Heithausen} {et~al.}(1998){Heithausen}, {Bensch}, {Stutzki},
  {Falgarone}, \& {Panis}}]{Heithausen98}
{Heithausen}, A., {Bensch}, F., {Stutzki}, J., {Falgarone}, E., \& {Panis},
  J.~F. 1998, \aap, 331, L65

\bibitem[{{Helou} {et~al.}(1988){Helou}, {Khan}, {Malek}, \&
  {Boehmer}}]{Helou1988}
{Helou}, G., {Khan}, I.~R., {Malek}, L., \& {Boehmer}, L. 1988, \apjs, 68, 151

\bibitem[{{Heyer} {et~al.}(2009){Heyer}, {Krawczyk}, {Duval}, \&
  {Jackson}}]{Heyer2008}
{Heyer}, M., {Krawczyk}, C., {Duval}, J., \& {Jackson}, J.~M. 2009, \apj, 699,
  1092

\bibitem[{{Heyer} \& {Terebey}(1998)}]{Heyer98}
{Heyer}, M.~H. \& {Terebey}, S. 1998, \apj, 502, 265

\bibitem[{{Hollenbach} \& {Tielens}(1999)}]{HollenbachTielens99}
{Hollenbach}, D.~J. \& {Tielens}, A.~G.~G.~M. 1999, Reviews of Modern Physics,
  71, 173

\bibitem[{{Imara} \& {Blitz}(2007)}]{Imara2007}
{Imara}, N. \& {Blitz}, L. 2007, \apj, 662, 969

\bibitem[{{Indebetouw} {et~al.}(2008){Indebetouw}, {Whitney}, {Kawamura},
  {Onishi}, {Meixner}, {Meade}, {Babler}, {Hora}, {Gordon}, {Engelbracht},
  {Block}, \& {Misselt}}]{Indebetouw2008}
{Indebetouw}, R., {Whitney}, B.~A., {Kawamura}, A., {Onishi}, T., {Meixner},
  M., {Meade}, M.~R., {Babler}, B.~L., {Hora}, J.~L., {Gordon}, K.,
  {Engelbracht}, C., {Block}, M., \& {Misselt}, K. 2008, \aj, 136, 1442

\bibitem[{{Israel}(2000)}]{Israel00}
{Israel}, F. 2000, in Molecular hydrogen in space, Cambridge, UK: Cambridge
  University Press, 2001. xix, 326 p.. Cambridge contemporary astrophysics.
  Edited by F. Combes, and G. Pineau des For{\^e}ts. ISBN 0521782244, p.293,
  ed. F.~{Combes} \& G.~{Pineau Des Forets}, 293

\bibitem[{{Israel}(1997)}]{Israel97}
{Israel}, F.~P. 1997, \aap, 328, 471

\bibitem[{{Israel} {et~al.}(2003){Israel}, {de Graauw}, {Johansson}, {Booth},
  {Boulanger}, {Garay}, {Kutner}, {Lequeux}, {Nyman}, \& {Rubio}}]{Israel2003}
{Israel}, F.~P., {de Graauw}, T., {Johansson}, L.~E.~B., {Booth}, R.~S.,
  {Boulanger}, F., {Garay}, G., {Kutner}, M.~L., {Lequeux}, J., {Nyman}, L.-A.,
  \& {Rubio}, M. 2003, \aap, 401, 99

\bibitem[{{Israel} {et~al.}(1996){Israel}, {Maloney}, {Geis}, {Herrmann},
  {Madden}, {Poglitsch}, \& {Stacey}}]{Israel96}
{Israel}, F.~P., {Maloney}, P.~R., {Geis}, N., {Herrmann}, F., {Madden}, S.~C.,
  {Poglitsch}, A., \& {Stacey}, G.~J. 1996, \apj, 465, 738

\bibitem[{{Johansson} {et~al.}(1998){Johansson}, {Greve}, {Booth}, {Boulanger},
  {Garay}, {de Graauw}, {Israel}, {Kutner}, {Lequeux}, {Murphy}, {Nyman}, \&
  {Rubio}}]{Johansson98}
{Johansson}, L.~E.~B., {Greve}, A., {Booth}, R.~S., {Boulanger}, F., {Garay},
  G., {de Graauw}, T., {Israel}, F.~P., {Kutner}, M.~L., {Lequeux}, J.,
  {Murphy}, D.~C., {Nyman}, L.-A., \& {Rubio}, M. 1998, \aap, 331, 857

\bibitem[{{Kim} {et~al.}(2007){Kim}, {Rosolowsky}, {Lee}, {Kim}, {Jung},
  {Dopita}, {Elmegreen}, {Freeman}, {Sault}, {Kesteven}, {McConnell}, \&
  {Chu}}]{Kim2007}
{Kim}, S., {Rosolowsky}, E., {Lee}, Y., {Kim}, Y., {Jung}, Y.~C., {Dopita},
  M.~A., {Elmegreen}, B.~G., {Freeman}, K.~C., {Sault}, R.~J., {Kesteven}, M.,
  {McConnell}, D., \& {Chu}, Y.-H. 2007, \apjs, 171, 419

\bibitem[{{Kramer} {et~al.}(2008){Kramer}, {Cubick}, {R{\"o}llig}, {Sun},
  {Yonekura}, {Aravena}, {Bensch}, {Bertoldi}, {Bronfman}, {Fujishita},
  {Fukui}, {Graf}, {Hitschfeld}, {Honingh}, {Ito}, {Jakob}, {Jacobs}, {Klein},
  {Koo}, {May}, {Miller}, {Miyamoto}, {Mizuno}, {Onishi}, {Park}, {Pineda},
  {Rabanus}, {Sasago}, {Schieder}, {Simon}, {Stutzki}, {Volgenau}, \&
  {Yamamoto}}]{Kramer2008}
{Kramer}, C., {Cubick}, M., {R{\"o}llig}, M., {Sun}, K., {Yonekura}, Y.,
  {Aravena}, M., {Bensch}, F., {Bertoldi}, F., {Bronfman}, L., {Fujishita}, M.,
  {Fukui}, Y., {Graf}, U.~U., {Hitschfeld}, M., {Honingh}, N., {Ito}, S.,
  {Jakob}, H., {Jacobs}, K., {Klein}, U., {Koo}, B.-C., {May}, J., {Miller},
  M., {Miyamoto}, Y., {Mizuno}, N., {Onishi}, T., {Park}, Y.-S., {Pineda},
  J.~L., {Rabanus}, D., {Sasago}, H., {Schieder}, R., {Simon}, R., {Stutzki},
  J., {Volgenau}, N., \& {Yamamoto}, H. 2008, \aap, 477, 547

\bibitem[{{Kramer} {et~al.}(1998){Kramer}, {Stutzki}, {Rohrig}, \&
  {Corneliussen}}]{Kramer98}
{Kramer}, C., {Stutzki}, J., {Rohrig}, R., \& {Corneliussen}, U. 1998, \aap,
  329, 249

\bibitem[{{Kutner} {et~al.}(1997){Kutner}, {Rubio}, {Booth}, {Boulanger}, {de
  Graauw}, {Garay}, {Israel}, {Johansson}, {Lequeux}, \& {Nyman}}]{Kutner97}
{Kutner}, M.~L., {Rubio}, M., {Booth}, R.~S., {Boulanger}, F., {de Graauw}, T.,
  {Garay}, G., {Israel}, F.~P., {Johansson}, L.~E.~B., {Lequeux}, J., \&
  {Nyman}, L.-A. 1997, \aaps, 122, 255

\bibitem[{{Ladd} {et~al.}(2005){Ladd}, {Purcell}, {Wong}, \&
  {Robertson}}]{Ladd05}
{Ladd}, N., {Purcell}, C., {Wong}, T., \& {Robertson}, S. 2005, Publications of
  the Astronomical Society of Australia, 22, 62

\bibitem[{{Larson}(1981)}]{Larson81}
{Larson}, R.~B. 1981, \mnras, 194, 809

\bibitem[{{Leroy} {et~al.}(2007){Leroy}, {Bolatto}, {Stanimirovic}, {Mizuno},
  {Israel}, \& {Bot}}]{Leroy2007}
{Leroy}, A., {Bolatto}, A., {Stanimirovic}, S., {Mizuno}, N., {Israel}, F., \&
  {Bot}, C. 2007, \apj, 658, 1027

\bibitem[{{Leroy} {et~al.}(2006){Leroy}, {Bolatto}, {Walter}, \&
  {Blitz}}]{Leroy2006}
{Leroy}, A., {Bolatto}, A., {Walter}, F., \& {Blitz}, L. 2006, \apj, 643, 825

\bibitem[{{Lisenfeld} \& {Ferrara}(1998)}]{Lisenfeld98}
{Lisenfeld}, U. \& {Ferrara}, A. 1998, \apj, 496, 145

\bibitem[{{MacLaren} {et~al.}(1988){MacLaren}, {Richardson}, \&
  {Wolfendale}}]{Maclaren88}
{MacLaren}, I., {Richardson}, K.~M., \& {Wolfendale}, A.~W. 1988, \apj, 333,
  821

\bibitem[{{Maloney} \& {Black}(1988)}]{MaloneyBlack88}
{Maloney}, P. \& {Black}, J.~H. 1988, \apj, 325, 389

\bibitem[{{May} {et~al.}(1997){May}, {Alvarez}, \& {Bronfman}}]{May1997}
{May}, J., {Alvarez}, H., \& {Bronfman}, L. 1997, \aap, 327, 325

\bibitem[{{McKee}(1989)}]{McKee1989}
{McKee}, C.~F. 1989, \apj, 345, 782

\bibitem[{{McKee} \& {Williams}(1997)}]{Mckee1997}
{McKee}, C.~F. \& {Williams}, J.~P. 1997, \apj, 476, 144

\bibitem[{{Meixner} {et~al.}(2006){Meixner}, {Gordon}, {Indebetouw}, {Hora},
  {Whitney}, {Blum}, {Reach}, {Bernard}, {Meade}, {Babler}, {Engelbracht},
  {For}, {Misselt}, {Vijh}, {Leitherer}, {Cohen}, {Churchwell}, {Boulanger},
  {Frogel}, {Fukui}, {Gallagher}, {Gorjian}, {Harris}, {Kelly}, {Kawamura},
  {Kim}, {Latter}, {Madden}, {Markwick-Kemper}, {Mizuno}, {Mizuno}, {Mould},
  {Nota}, {Oey}, {Olsen}, {Onishi}, {Paladini}, {Panagia}, {Perez-Gonzalez},
  {Shibai}, {Sato}, {Smith}, {Staveley-Smith}, {Tielens}, {Ueta}, {Dyk},
  {Volk}, {Werner}, \& {Zaritsky}}]{Meixner06}
{Meixner}, M., {Gordon}, K.~D., {Indebetouw}, R., {Hora}, J.~L., {Whitney}, B.,
  {Blum}, R., {Reach}, W., {Bernard}, J.-P., {Meade}, M., {Babler}, B.,
  {Engelbracht}, C.~W., {For}, B.-Q., {Misselt}, K., {Vijh}, U., {Leitherer},
  C., {Cohen}, M., {Churchwell}, E.~B., {Boulanger}, F., {Frogel}, J.~A.,
  {Fukui}, Y., {Gallagher}, J., {Gorjian}, V., {Harris}, J., {Kelly}, D.,
  {Kawamura}, A., {Kim}, S., {Latter}, W.~B., {Madden}, S., {Markwick-Kemper},
  C., {Mizuno}, A., {Mizuno}, N., {Mould}, J., {Nota}, A., {Oey}, M.~S.,
  {Olsen}, K., {Onishi}, T., {Paladini}, R., {Panagia}, N., {Perez-Gonzalez},
  P., {Shibai}, H., {Sato}, S., {Smith}, L., {Staveley-Smith}, L., {Tielens},
  A.~G.~G.~M., {Ueta}, T., {Dyk}, S.~V., {Volk}, K., {Werner}, M., \&
  {Zaritsky}, D. 2006, \aj, 132, 2268

\bibitem[{{Minamidani} {et~al.}(2007){Minamidani}, {Mizuno}, {Mizuno},
  {Kawamura}, {Onishi}, {Hasegawa}, {Tatematsu}, {Ikeda}, {Moriguchi},
  {Yamaguchi}, {Ott}, {Wong}, {Muller}, {Pineda}, {Hughes}, {Staveley-Smith},
  {Klein}, {Mizuno}, {Nikoli{\'c}}, {Booth}, {Heikkil{\"a}}, {Nyman}, {Lerner},
  {Garay}, {Kim}, {Fujishita}, {Kawase}, {Rubio}, \& {Fukui}}]{Minamidani2007}
{Minamidani}, T., {Mizuno}, N., {Mizuno}, Y., {Kawamura}, A., {Onishi}, T.,
  {Hasegawa}, T., {Tatematsu}, K., {Ikeda}, M., {Moriguchi}, Y., {Yamaguchi},
  N., {Ott}, J., {Wong}, T., {Muller}, E., {Pineda}, J.~L., {Hughes}, A.,
  {Staveley-Smith}, L., {Klein}, U., {Mizuno}, A., {Nikoli{\'c}}, S., {Booth},
  R.~S., {Heikkil{\"a}}, A., {Nyman}, L.~., {Lerner}, M., {Garay}, G., {Kim},
  S., {Fujishita}, M., {Kawase}, T., {Rubio}, M., \& {Fukui}, Y. 2007, ArXiv
  e-prints, 710

\bibitem[{{Mizuno} {et~al.}(2001{\natexlab{a}}){Mizuno}, {Yamaguchi}, {Mizuno},
  {Rubio}, {Abe}, {Saito}, {Onishi}, {Yonekura}, {Yamaguchi}, {Ogawa}, \&
  {Fukui}}]{Mizuno01}
{Mizuno}, N., {Yamaguchi}, R., {Mizuno}, A., {Rubio}, M., {Abe}, R., {Saito},
  H., {Onishi}, T., {Yonekura}, Y., {Yamaguchi}, N., {Ogawa}, H., \& {Fukui},
  Y. 2001{\natexlab{a}}, \pasj, 53, 971

\bibitem[{{Mizuno} {et~al.}(2001{\natexlab{b}}){Mizuno}, {Yamaguchi}, {Mizuno},
  {Rubio}, {Abe}, {Saito}, {Onishi}, {Yonekura}, {Yamaguchi}, {Ogawa}, \&
  {Fukui}}]{Mizuno2001}
---. 2001{\natexlab{b}}, \pasj, 53, 971

\bibitem[{{Mookerjea} {et~al.}(2006){Mookerjea}, {Kramer}, {R{\"o}llig}, \&
  {Masur}}]{Mookerjea2006}
{Mookerjea}, B., {Kramer}, C., {R{\"o}llig}, M., \& {Masur}, M. 2006, \aap,
  456, 235

\bibitem[{{Myers}(1983)}]{Myers83}
{Myers}, P.~C. 1983, \apj, 270, 105

\bibitem[{{Nakagawa} {et~al.}(1998){Nakagawa}, {Yui}, {Doi}, {Okuda}, {Shibai},
  {Mochizuki}, {Nishimura}, \& {Low}}]{Nakagawa1998}
{Nakagawa}, T., {Yui}, Y.~Y., {Doi}, Y., {Okuda}, H., {Shibai}, H.,
  {Mochizuki}, K., {Nishimura}, T., \& {Low}, F.~J. 1998, \apjs, 115, 259

\bibitem[{{Ott} {et~al.}(2008){Ott}, {Wong}, {Pineda}, {Hughes}, {Muller},
  {Li}, {Wang}, {Staveley-Smith}, {Fukui}, {Wei{\ss}}, {Henkel}, \&
  {Klein}}]{Ott2008}
{Ott}, J., {Wong}, T., {Pineda}, J.~L., {Hughes}, A., {Muller}, E., {Li},
  Z.-Y., {Wang}, M., {Staveley-Smith}, L., {Fukui}, Y., {Wei{\ss}}, A.,
  {Henkel}, C., \& {Klein}, U. 2008, Publications of the Astronomical Society
  of Australia, 25, 129

\bibitem[{{Pak} {et~al.}(1998){Pak}, {Jaffe}, {van Dishoeck}, {Johansson}, \&
  {Booth}}]{Pak1998}
{Pak}, S., {Jaffe}, D.~T., {van Dishoeck}, E.~F., {Johansson}, L.~E.~B., \&
  {Booth}, R.~S. 1998, \apj, 498, 735

\bibitem[{{Pineda} {et~al.}(2008){Pineda}, {Mizuno}, {Stutzki}, {Cubick},
  {Aravena}, {Bensch}, {Bertoldi}, {Bronfman}, {Fujishita}, {Graf},
  {Hitschfeld}, {Honingh}, {Jakob}, {Jacobs}, {Kawamura}, {Klein}, {Kramer},
  {May}, {Miller}, {Mizuno}, {M{\"u}ller}, {Onishi}, {Ossenkopf}, {Rabanus},
  {R{\"o}llig}, {Rubio}, {Sasago}, {Schieder}, {Simon}, {Sun}, {Volgenau},
  {Yamamoto}, \& {Fukui}}]{Pineda2008}
{Pineda}, J.~L., {Mizuno}, N., {Stutzki}, J., {Cubick}, M., {Aravena}, M.,
  {Bensch}, F., {Bertoldi}, F., {Bronfman}, L., {Fujishita}, K., {Graf}, U.~U.,
  {Hitschfeld}, M., {Honingh}, N., {Jakob}, H., {Jacobs}, K., {Kawamura}, A.,
  {Klein}, U., {Kramer}, C., {May}, J., {Miller}, M., {Mizuno}, Y.,
  {M{\"u}ller}, P., {Onishi}, T., {Ossenkopf}, V., {Rabanus}, D., {R{\"o}llig},
  M., {Rubio}, M., {Sasago}, H., {Schieder}, R., {Simon}, R., {Sun}, K.,
  {Volgenau}, N., {Yamamoto}, H., \& {Fukui}, Y. 2008, \aap, 482, 197

\bibitem[{{Rosolowsky}(2005)}]{Rosolowsky2005}
{Rosolowsky}, E. 2005, \pasp, 117, 1403

\bibitem[{{Rosolowsky} \& {Leroy}(2006)}]{Rosolowsky2006}
{Rosolowsky}, E. \& {Leroy}, A. 2006, \pasp, 118, 590

\bibitem[{{Rubio} {et~al.}(2004){Rubio}, {Boulanger}, {Rantakyro}, \&
  {Contursi}}]{Rubio2004}
{Rubio}, M., {Boulanger}, F., {Rantakyro}, F., \& {Contursi}, A. 2004, \aap,
  425, L1

\bibitem[{{Rubio} {et~al.}(1991){Rubio}, {Garay}, {Montani}, \&
  {Thaddeus}}]{Rubio91}
{Rubio}, M., {Garay}, G., {Montani}, J., \& {Thaddeus}, P. 1991, \apj, 368, 173

\bibitem[{{Rubio} {et~al.}(1993){Rubio}, {Lequeux}, \& {Boulanger}}]{Rubio93}
{Rubio}, M., {Lequeux}, J., \& {Boulanger}, F. 1993, \aap, 271, 9

\bibitem[{{Sakamoto}(1996)}]{Sakamoto96}
{Sakamoto}, S. 1996, \apj, 462, 215

\bibitem[{{Schwering}(1989)}]{Schwering1989}
{Schwering}, P.~B.~W. 1989, \aaps, 79, 105

\bibitem[{{Sheth} {et~al.}(2008){Sheth}, {Vogel}, {Wilson}, \&
  {Dame}}]{Sheth2008}
{Sheth}, K., {Vogel}, S.~N., {Wilson}, C.~D., \& {Dame}, T.~M. 2008, \apj, 675,
  330

\bibitem[{{Smith} {et~al.}(1987){Smith}, {Cornett}, \& {Hill}}]{Smith1987}
{Smith}, A.~M., {Cornett}, R.~H., \& {Hill}, R.~S. 1987, \apj, 320, 609

\bibitem[{{Solomon} {et~al.}(1987){Solomon}, {Rivolo}, {Barrett}, \&
  {Yahil}}]{Solomon87}
{Solomon}, P.~M., {Rivolo}, A.~R., {Barrett}, J., \& {Yahil}, A. 1987, \apj,
  319, 730

\bibitem[{{Stanimirovi{\'c}} \& {Lazarian}(2001)}]{Staminirovic2001}
{Stanimirovi{\'c}}, S. \& {Lazarian}, A. 2001, \apjl, 551, L53

\bibitem[{{Stanimirovi{\'c}} {et~al.}(2000){Stanimirovi{\'c}},
  {Staveley-Smith}, {van der Hulst}, {Bontekoe}, {Kester}, \&
  {Jones}}]{Staminirivic2000}
{Stanimirovi{\'c}}, S., {Staveley-Smith}, L., {van der Hulst}, J.~M.,
  {Bontekoe}, T.~R., {Kester}, D.~J.~M., \& {Jones}, P.~A. 2000, \mnras, 315,
  791

\bibitem[{{Stutzki} \& {G\"usten}(1990)}]{StutzkiGuesten90}
{Stutzki}, J. \& {G\"usten}, R. 1990, \apj, 356, 513

\bibitem[{{Stutzki} {et~al.}(1988){Stutzki}, {Stacey}, {Genzel}, {Harris},
  {Jaffe}, \& {Lugten}}]{Stutzki88}
{Stutzki}, J., {Stacey}, G.~J., {Genzel}, R., {Harris}, A.~I., {Jaffe}, D.~T.,
  \& {Lugten}, J.~B. 1988, \apj, 332, 379

\bibitem[{{Tielens} \& {Hollenbach}(1985)}]{TielensHollenbach85}
{Tielens}, A.~G.~G.~M. \& {Hollenbach}, D. 1985, \apj, 291, 722

\bibitem[{{van der Marel} \& {Cioni}(2001)}]{vanderMarel01}
{van der Marel}, R.~P. \& {Cioni}, M.-R.~L. 2001, \aj, 122, 1807

\bibitem[{{van Dishoeck} \& {Black}(1988)}]{vanDishBlack88}
{van Dishoeck}, E.~F. \& {Black}, J.~H. 1988, \apj, 334, 771

\bibitem[{{Verter} \& {Hodge}(1995)}]{Verter95}
{Verter}, F. \& {Hodge}, P. 1995, \apj, 446, 616

\bibitem[{{Walmsley} {et~al.}(2000){Walmsley}, {Natta}, {Oliva}, \&
  {Testi}}]{Walmsley00}
{Walmsley}, C.~M., {Natta}, A., {Oliva}, E., \& {Testi}, L. 2000, \aap, 364,
  301

\bibitem[{{Wei{\ss}} {et~al.}(2001){Wei{\ss}}, {Neininger}, {H{\"u}ttemeister},
  \& {Klein}}]{Weiss01}
{Wei{\ss}}, A., {Neininger}, N., {H{\"u}ttemeister}, S., \& {Klein}, U. 2001,
  \aap, 365, 571

\bibitem[{{Westerlund}(1997)}]{Westerlund97}
{Westerlund}, B.~E. 1997, {The Magellanic Clouds} (New York: Cambridge Univ.
  Press)

\bibitem[{{Williams} {et~al.}(1994){Williams}, {de Geus}, \&
  {Blitz}}]{Williams94}
{Williams}, J.~P., {de Geus}, E.~J., \& {Blitz}, L. 1994, \apj, 428, 693

\bibitem[{{Wilson}(1995)}]{Wilson95}
{Wilson}, C.~D. 1995, \apjl, 448, L97+

\bibitem[{{Young Owl} {et~al.}(2000){Young Owl}, {Meixner}, {Wolfire},
  {Tielens}, \& {Tauber}}]{YoungOwl00}
{Young Owl}, R.~C., {Meixner}, M.~M., {Wolfire}, M., {Tielens}, A.~G.~G.~M., \&
  {Tauber}, J. 2000, \apj, 540, 886

\end{thebibliography}
\bibliographystyle{apj}

\clearpage

\end{document}